\documentclass[useAMS,usenatbib,onecolumn]{mn2e}
\usepackage{epsfig}
\usepackage{graphicx}
\usepackage{amsmath}  
\newcommand{\etal}{{et~al.}~}
\newcommand{\lsim}{\,\lower2truept\hbox{${<\atop\hbox{\raise4truept\hbox{$\sim$}}}$}\,}
\newcommand{\gsim}{\,\lower2truept\hbox{${>\atop\hbox{\raise4truept\hbox{$\sim$}}}$}\,}
\newcommand{\beq}{\begin{equation}}
\newcommand{\eeq}{\end{equation}}

\newcommand{\WMAP}{$WMAP$}
\newcommand{\fastica}{{\sc{fastica}}}

\def\aa{{\sl Astron.\ \&\ Astrophys.\ }}
\def\aas{{\sl Astron. \& Astrophys.\ Suppl.\ }}
\def\apj{{\sl Astrophys.\ J.\ }}

\def\apjs{{\sl Astrophys.\ J.\ Supp.\ }}

\def\ieeespl{{\sl IEEE\ Signal\ Processing\ Lett.\ }}

\def\mnras{{\sl MNRAS\ }}

\def\nn{{\sl Neural Networks\ }}

\def\prd{{\sl Phys.\ Rev.\ D\ }}

\title[Foreground analysis of the WMAP 5yr data using \fastica]
{New insights into foreground analysis of the WMAP five-year data using \fastica}

\author[Bottino \etal]{M. Bottino$^{1}\footnote{E-mail:bottino@mpa-garching.mpg.de}$,
A.~J. Banday$^{2,1}$, D. Maino$^3$\\
$^{1}$ Max-Planck Institute f{\"u}r Astrophysik, Karl-Schwarzschild Str. 1, D-85748, Garching, Germany\\
$^{2}$ Centre d'Etude Spatiale des Rayonnements, 9, av du Colonel Roche, BP 44346, 31028 Toulouse Cedex 4, France\\
$^{3}$ Dipartimento di Fisica, Universit\`a di Milano, Via Celoria 16, I-20133, Milano, Italy\\
}

\pagerange{\pageref{firstpage}--\pageref{lastpage}}
\pubyear{2006}

\begin{document}

\maketitle
\label{firstpage}

\begin{abstract}

In this paper, we present
a foreground analysis of the \WMAP\ 5-year data using the {\fastica}
algorithm, improving on the treatment of the \WMAP\ 3-year data in \citet{bottino_etal_2008}.

We determine coupling coefficients between the \WMAP\ data and
templates commonly used to trace the dominant foreground emissions, 
and use them to study the spectral properties of the diffuse emissions
and subsequently to clean the data.
We confirm the dependence of the values of these scaling factors on the extension of the mask used 
in the analysis and we again demonstrate some anomalies 
when the $Kp0$ mask is adopted that remain unexplained.

We also revisit the nature of the free-free spectrum with the emphasis 
on attempting to confirm or otherwise the spectral feature 
claimed in \citet{dobler_etal_2008b} and explained in terms of
spinning dust emission in the warm ionised medium. With the
application of different Galactic cuts, the index is always flatter 
than the canonical value of $2.14$ except for the $Kp0$ mask which is 
steeper. Irrespective of this, we can not
confirm the presence of any feature in the free-free
spectrum. 

We experiment with a more extensive approach to the cleaning of the data,
introduced in connection with the iterative application of {\fastica}.
We confirm the presence of a residual foreground whose spatial
distribution is concentrated along the Galactic plane, with pronounced
emission near the Galactic center. This is consistent with the
\WMAP\ haze detected in \citet{finkbeiner_2004}.

Finally, we attempted to perform the same analysis on full-sky
maps. The code returns good results even for those regions where the
cross-talk among the components is high.  However, slightly better
results in terms of the possibility of reconstructing a full-sky CMB
map, are achieved with a simultaneous analysis of both the five \WMAP\ maps
and foreground templates.  Nonetheless, some residuals are still
present and detected in terms of an excess in the CMB power spectrum,
on small angular scales. Therefore, a minimal mask for the brightest
regions of the plane is necessary, and has been defined.

\end{abstract}

\footnotetext{E-mail: bottino@mpa-garching.mpg.de}

\section{Introduction}
\label{intro}
The most recent release of five years of observations from the 
\WMAP\footnote{\emph{Wilkinson Microwave Anisotropy Probe}} satellite
again allows a quantitative improvement 
in studies of the Cosmic Microwave Background (CMB) for cosmological purposes.
However, such an improvement requires an analogous refinement in both
our understanding of local astrophysical foregrounds and in the methods
employed for the separation of these components from the CMB emission.

The most evident source of foreground contamination in the \WMAP\ data
on at least large-angular scales is associated with the radiation
produced by our own Galaxy -- due to synchrotron emission, free-free
emission (or \emph{thermal bremsstrahlung}) and thermal dust emission.
The presence of an additional component is now also almost universally
accepted -- this emission is strongly correlated with dust emission at
microwave wavelengths but with a spectral dependence that is
inconsistent with the basic thermal mechanism. It is generally
referred to as \emph{anomalous dust emission} since its nature is
still not unambiguously identified.  Finally, a subject of ongoing
debate is the so-called \lq \WMAP\ haze' which
\citet{dobler_etal_2008a} attribute to \emph{hard} synchrotron
emission distributed around the Galactic Center, again with uncertain
physical origin.

Many different techniques have been developed in order to clean 
the CMB data and subsequently study the foreground emission components.
An inexhaustive selection of these includes: linear combination methods 
\citep{bennett_etal_2003,tegmark_etal_2003,park_etal_2007,kim_etal_2008,delabrouille_etal_2008}, 
Gibbs sampling \citep{eriksen_etal_2008},  the Maximum Entropy Method (MEM) \citep{bennett_etal_2003,hinshaw_etal_2007},
WIFIT \citep{hansen_etal_2006}, 
Correlated Component Analysis (CCA) \citep{bonaldi_etal_2007}, and 
SMICA \citep{delabrouille_etal_2003,patanchon_etal_2005,cardoso_etal_2008}.  
In addition, 
an implementation of the \emph{Independent Component Analysis}
approach \citep{hyvarinen_1999,hyvarinen_oja_2000} referred to as
\fastica\ has been successfully applied to the problem
in \citet{maino_etal_2002} and
\citet{baccigalupi_etal_2004}. 

The aim of the current work is to undertake the foreground analysis of
the \WMAP\ 5-year data with this algorithm, as previously applied
to the \WMAP\ 3-year data \citep{bottino_etal_2008}. As before, we
focus on the diffuse Galactic foreground components
and their spectral and spatial properties.  Specifically, we first
apply \fastica\ to a set of data comprising one of the \WMAP\ 
frequency maps together with three standard templates of the diffuse
Galactic emission, as described in section~\ref{data} and section~\ref{data_analysis}.
We derive coupling coefficients for each frequency that, when used to
scale the template amplitudes, are interpreted as the amount of
foreground contamination for each foreground component at that
frequency.  Then, we iteratively apply the algorithm to the set of
data, cleaned according to these scale factors.  This second step of
the analysis allows us to identify physical residuals that are not
uniquely identified with one of the three templates.

Novel aspects of the analysis introduced in this paper include:

\begin{itemize}
  
  \item consideration of the new KQ85 and KQ75 masks introduced by the
    \WMAP\ science team with the 5-year data release

  \item an investigation into the connection between the mask applied to derive the 
    coupling coefficients and  the residuals resulting from the iterative multi-frequency
    analysis when these coefficients are applied, itself as a function of applied mask

  \item an attempt to understand the physical nature of the residuals revealed by the 
    iterative analysis.

  \item the definition of a minimal mask that supports accurate foreground removal

\end{itemize}

A general summary of the paper is as follows.  After a brief review in
section~\ref{ica_templates} on how we use {\fastica} to perform the
analysis, in section~\ref{data} we describe the data and templates
adopted, pointing out the main differences with respect to previous
work. Then, in section~\ref{simulations} we calibrate the performance of the
method using Monte Carlo simulations and by comparison to a simpler
$\chi^2$-based method.  In section \ref{data_analysis}, we present the
results of the template fits, and study the spectral behaviour of the
foregrounds with particular emphasis on the free-free emission.  The
scaled templates are then used to clean the data, the resultant
properties of which are investigated in section \ref{cleaned_data}.
In subsequent sections the iterative application of {\fastica} then attempts to improve the
removal of foreground contamination from the data.  Finally, we
experiment with full-sky analysis, which benefits from the
simultaneous analysis of multi-frequency data and templates.
Section~\ref{discussion} summarises our main conclusions.

\section{\fastica\ AND ITS USE FOR FOREGROUND COMPONENT STUDIES}
\label{ica_templates}

We apply \fastica\ to the \WMAP\ 5-year data as described in
\citet{bottino_etal_2008}.
We refer explicitly to that paper and to \citet{maino_etal_2007} 
for details about how the code works, 
particularly in the context of template fitting.
Nevertheless, we remind the reader that the algorithm
determines a linear transformation of the input maps
by seeking the maxima of their  \emph{neg-entropy}. 
In the implementation used here, this quantity is generally approximated by three non-linear 
functions, i.e. $p(u) = u^3$, $t(u) = 
{\rm tanh}(u)$ and $g(u) = u\, {\rm exp}(-u^2)$ where $u$ are the
principal component projected data \citep{hyvarinen_1999,hyvarinen_oja_2000}.
Generally, $p$, 
which corresponds to the kurtosis, should be used for sub-Gaussian 
components but it is strongly sensitive to outliers in the distributions; 
$g$ is for super-Gaussian signals while $t$ is a general purpose function.
The suitability of these functions for our purposes are described in
section~\ref{simulations}.

As before, we also compare our analysis to the simpler $\chi^2$-based
template fitting scheme, commonly used in the field, as a convenient
point of reference for our results (see \citet{bottino_etal_2008} for
details).  To be consistent with the {\fastica} analysis, we did not
impose any constraints on the data, although this is possible as
described in \citet{hinshaw_etal_2007} and \citet{gold_etal_2009}.

The computed scaling factors are then used to clean the \WMAP\ data
from the estimated Galactic contaminations. The resultant maps still
show evidence of residuals due to the imperfect correlation of the true
foreground emission at microwave frequencies with the adopted
templates, arising in part from the assumption of a fixed spectral index for each
physical foreground component over the analysed sky coverage.
As shown in the analysis of the \WMAP\ 3-year data, a subsequent
iterative application of the \fastica\ algorithm to this
multi-frequency pre-cleaned microwave data can
yield an improved estimate of the CMB sky, plus component maps that
represent foreground residuals.

\section{DATA USED IN THE ANALYSIS}
\label{data}

Our primary data set consists of the \WMAP\ 5-year maps 
improved in sensitivity and precision 
as described in \citet{hinshaw_etal_2009},
and available for download on the LAMBDA
website\footnote{\emph{ Legacy Archive for Microwave Background Data
Analysis} -- http://lambda.gsfc.nasa.gov/.} 
in a  HEALPix\footnote{http://healpix.jpl.nasa.gov.} 
pixelisation scheme, with a pixel resolution parameter of $N_{side}=512$.

In our analysis, we assume that the data can be well described by a
superposition of the CMB anisotropy plus foreground emission traced by
three templates. These templates, suitably scaled to each frequency,
are considered to represent four different physical components:
synchrotron, free-free and thermal dust emission, together with the
anomalous dust component -- a dust correlated emission whose detailed
nature is not well known, except for the fact that its contribution
increases with decreasing frequency below $\sim\ 61$ GHz.  Evidence of
a turn-over in the spectrum below $\sim\ 15$ GHz may be evidence that
the emission arises from so-called \lq spinning dust'
\citep{draine_etal_1998}. Irrespective of its detailed physical
origins, its existence is generally accepted and observed in many
independent analyses.

The templates are based on maps produced from independent observations
of the sky at frequencies where only one emission mechanism dominates. 
In practise, we retain the set of templates used for the \WMAP\ 3-year
analysis, in order to be able to relate all
changes in the analysis results to the improvements in the \WMAP\ data. 
Therefore, we used the 408 MHz radio
continuum all-sky map of \citet{haslam_etal_1982} which is
dominated by synchrotron emission away from the Galactic plane. 
The all-sky H$\alpha$-map, produced by \citet{finkbeiner_2003} by assembling
data from several surveys, is the template utilised for the free-free emission.
The map was not corrected for dust absorption, assuming it to be negligible as
suggested by \citet{banday_etal_2003}. Nevertheless, we tested the dependence of our
results to this correction applied on the template. 
Moreover, we have also tested an alternative H$\alpha$-map
provided by \citet{dickinson_etal_2003}. This has demonstrated some
differences in correlation properties with the \WMAP\ data in previous
analyses \citep{davies_etal_2006}, but we do not find any sensitivity
to the choice of  H$\alpha$ template in our work.

Finally, we used the model for thermal dust emission at 94 GHz developed by \citet{finkbeiner_etal_1999} (FDS) 
based on the \emph{COBE}-DIRBE 100 and 240 $\mu$m  maps
as calibrated by the \emph{COBE}-FIRAS spectral data in the range 0.14 to 3.0~mm. 
This template was also used for the anomalous dust emission.
More detailed descriptions of these foregrounds and the corresponding
templates may be found, for example, in \citet{davies_etal_2006}. 
However, we note that the synchrotron emission (in antenna
temperature) is generally described by a power-law spectrum
$T_{s}(\nu)\sim \nu^{-\beta_s}$, similarly the free-free by $T_{ff}\sim
\nu^{-\beta_{ff}}$, and the thermal dust approximated as $T_{d}\sim
\nu^{\beta_{d}}$ over the \WMAP\ range of frequencies.

Of course, as noted in our previous paper, the choice of templates may
not necessarily be the optimal one, although the maps utilised are
those most frequently used in the literature. 
Indeed,
\citet{ysard_etal_2009} claim to have found evidence that the anomalous
emission is preferentially correlated with the IRAS 12 $\mu$m maps.
However, in the specific
case of the synchrotron emission, the \WMAP\ team have suggested that
the 408 MHz survey is significantly in error at microwave wavelengths,
and proposed that a more realistic description of the synchrotron sky
is provided by the difference of their K- and Ka-band data
\citep{hinshaw_etal_2007}. We therefore also include this template as
an alternative in our analysis.

These templates are used to fit the spectral behaviour of the \WMAP\ 
data at each frequency of observation from $\sim$23~GHz (K-band) up to
$\sim$94~GHz (W-band). Where multiple samples are available at a given
frequency, we have coadded them into a single sky map using
uniform and equal weights.  As before, we performed our analysis on
sky maps convolved from their original resolution to an effective
$1^{\circ}$ Gaussian beam.  Finally, we converted the \WMAP\ data from
thermodynamic temperature to brightness (antenna) temperature units,
which is more natural for an analysis of foreground spectral
behaviour.

Since regions close to the Galactic plane are the most seriously
contaminated by foregrounds and the spectral and spatial nature of the
integrated emission is complex, they should generally be excluded from
analysis. Indeed the \WMAP\ science team itself has proposed two new
different masks than the previously used $Kp2$ and $Kp0$ cuts. They
refer to the new masks as $KQ85$ and $KQ75$ since they are produced
from the maps at the K and Q band and they exclude respectively $15\%$
and $25\%$ of the sky \citep{hinshaw_etal_2009}. We employed all 4
masks together with a mask for the point sources, without any
modification of the angular resolution.  However, in the latter part
of the paper we also experiment with a full-sky analysis as shown in
section \ref{fastica_experiments}.

\section{MONTE CARLO SIMULATIONS}
\label{simulations}

\begin{figure}
\begin{centering}
\includegraphics[angle=90,width=1.\textwidth]{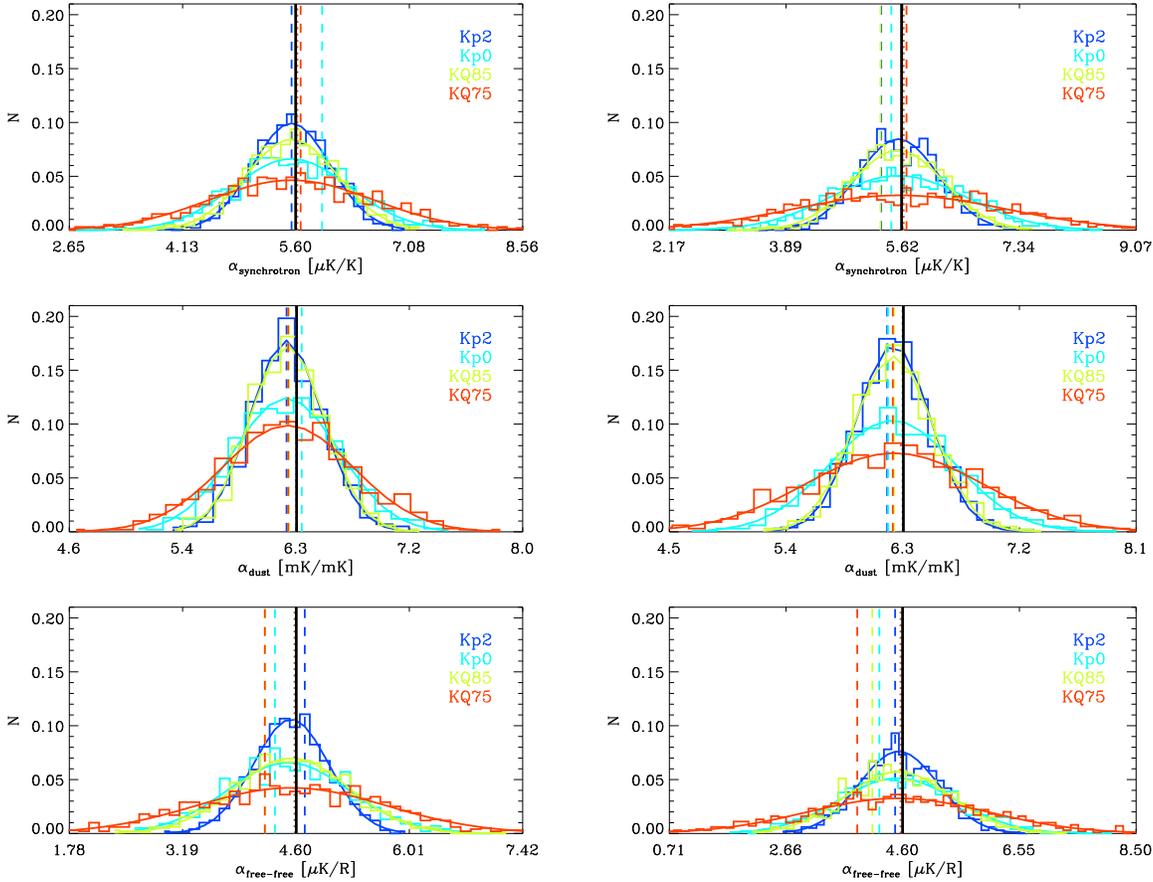}
\caption{Histogram of the coupling coefficient distributions at K-band obtained from 
the simulations using the $p$ (left) and $g$ (right) functions and all the masks. Different color 
are associated to different masks in order to compare the results: blue for $Kp2$, cyan for $Kp0$, 
green for $KQ85$ and red for $KQ75$. The dotted and dashed line show respectively the mean and the 
mode of the distributions, to be compared with the expected values (black line). For all the 
Galactic components except the dust emission, there is a strong dependence of the distribution 
width on the extension of the plane cut. This is particularly visible for the free-free emission 
where the scaling factors obtained with the KQ75 mask are characterized by a very flat distribution.  
  }
\label{graph_hist_nch_1_allmasks}
\end{centering}
\end{figure}

Monte Carlo simulations are used in this context for several reasons.
First, they 
provide useful criteria and figures-of-merit that aid the selection of
the non-linear function ($p$, $g$ or $t$) used by the \fastica\ algorithm to implement
the component separation.  Second, they provide a method to evaluate
the level of correlation with and among the foregrounds.  Finally we
used these simulations to compute the uncertainties for the scaling
factors presented in later sections of the paper.

We find that essentially all of the previous results from the 3-year analysis
are confirmed. 
The reconstruction of the input coefficients is satisfactory for both
the $p$ and $g$ functions, which provide similar results.  Indeed, the
distribution of the coefficients is a symmetric Gaussian function in
both cases: the mean value is very close to the expected one and
the mode differs only slightly. However, the $g$ function shows
generally a broader distribution compared to that for $p$, most likely
as a consequence of the fact that they are sensitive to different
statistical features of the spatial distribution of the sky radiation.
For the same reason the $t$-function remains inappropriate for
template fitting: the statistical distributions of the returned
coupling coefficients are asymmetric and highly biased with respect to
the input values.

The distribution of the scaling factors generally depends on the mask
used in the analysis (see figure \ref{graph_hist_nch_1_allmasks}): the
more extensive the cut is, the larger is the dispersion of the values
around the mean. It is particularly evident for the free-free emission
where the $KQ75$ mask is associated with the largest distribution,
though it is still symmetric and not biased.  We also find that the
anti-correlation between the synchrotron and dust coefficients
decreases for larger masks, as is the case for the correlation between
the K-Ka template and the free-free.

The results from the simple $\chi^2$ analysis indicate that the
uncertainties for the synchrotron component are smaller than found
using the {\fastica} method, while they are larger for the other
components.
This method dependent cross-talk likely reflects the different
statistical measures of the data employed by the {\fastica} and
$\chi^2$ analyses.

\section{COUPLING COEFFICIENTS BETWEEN DATA AND TEMPLATES}
\label{data_analysis}

As noted above, the \fastica\ component separation allows us to compute 
coupling coefficients between the data at each frequency of observation, and 
the three templates of the Galactic foreground components.
We calculated these values using the $p$ and the $g$ non-linear functions and the 
four masks provided by the \WMAP\ science team. Then, we compared the results with those 
from the simple $\chi^2$ analysis.
In table \ref{table_haslam_coeff} and \ref{table_k-ka_coeff}, we report the results
obtained with two different sets of templates, where either the Haslam 
or the K-Ka map was used as synchrotron model.
In the same tables, we show also the results from \citet{bennett_etal_2003}, 
\citet{hinshaw_etal_2007} and  \citet{gold_etal_2009}: these values are the 
\WMAP\ results respectively for the 1-, 3- and and 5-year foreground analysis
corrections, with the $Kp2$ or $KQ85$ masks. 
In all three cases, some constraints were placed on the spectral behaviour of the
components. 
In the first case the values 
were derived by fixing the synchrotron spectral index at a
value of 2.7, and the free-free to 2.15 for frequencies above Q-band. 
In the latter cases, instead, they explicitly take into account
the free-free signal present in the K-Ka template, and
impose constraints on the free-free ($\beta_{ff}=2.14$) and thermal
dust ($\beta_{d}=2$) spectral indices. 
However, we did not consider any constraints on the coefficients.

\begin{table}
\begin{center}
\begin{tabular}{l cccccc } 
\hline
\hline
& \multicolumn{2}{c}{\bf{synchrotron}} &  \multicolumn{2}{c}{\bf{dust}} &  \multicolumn{2}{c}{\bf{free-free}}  \\ 
\hline
&$Kp2$&$Kp0$& $Kp2$&$Kp0$&$Kp2$&$Kp0$\\
\hline
&\multicolumn{6}{c}{\bf{{\fastica} - function p }}\\
\hline
$K$&$6.20\pm0.49$&$5.47\pm0.75$&$5.88\pm0.29$&$4.96\pm0.40$&$8.57\pm0.47$&$6.66\pm0.75$\\
$Ka$&$1.91\pm0.49$&$1.70\pm0.74$&$2.04\pm0.28$&$1.29\pm0.38$&$4.35\pm0.46$&$2.74\pm0.74$\\
$Q$&$0.96\pm0.48$&$0.87\pm0.74$&$1.09\pm0.28$&$0.44\pm0.37$&$2.91\pm0.46$&$1.34\pm0.73$\\
$V$&$0.26\pm0.46$&$0.24\pm0.70$&$0.61\pm0.26$&$0.05\pm0.35$&$1.34\pm0.43$&$-0.07\pm0.70$\\
$W$&$0.06\pm0.40$&$0.07\pm0.61$&$0.97\pm0.23$&$0.46\pm0.31$&$0.64\pm0.38$&$-0.52\pm0.61$\\
\hline
&\multicolumn{6}{c}{\bf{{\fastica} - function g }}\\
\hline
$K$&$6.42\pm0.58$&$5.63\pm0.95$&$6.14\pm0.30$&$5.76\pm0.48$&$8.48\pm0.65$&$6.52\pm0.97$\\
$Ka$&$2.04\pm0.56$&$1.91\pm0.94$&$2.10\pm0.29$&$1.68\pm0.47$&$4.26\pm0.64$&$2.80\pm0.95$\\
$Q$&$1.10\pm0.56$&$1.12\pm0.92$&$1.11\pm0.29$&$0.71\pm0.47$&$2.79\pm0.63$&$1.45\pm0.94$\\
$V$&$0.40\pm0.53$&$0.54\pm0.87$&$0.60\pm0.27$&$0.24\pm0.44$&$1.21\pm0.60$&$0.03\pm0.89$\\
$W$&$0.19\pm0.46$&$0.33\pm0.77$&$0.95\pm0.24$&$0.61\pm0.39$&$0.56\pm0.53$&$-0.41\pm0.78$\\
\hline
&\multicolumn{6}{c}{\bf{$\chi^2$ analysis}} \\
\hline
$K$&$5.98\pm0.42$&$5.42\pm0.57$&$6.44\pm0.31$&$6.34\pm0.55$&$7.93\pm0.74$&$5.68\pm1.20$\\
$Ka$&$1.84\pm0.42$&$1.62\pm0.56$&$2.26\pm0.30$&$2.12\pm0.54$&$3.77\pm0.73$&$2.11\pm1.18$\\
$Q$&$0.92\pm0.41$&$0.79\pm0.56$&$1.24\pm0.30$&$1.12\pm0.53$&$2.31\pm0.72$&$0.80\pm1.17$\\
$V$&$0.24\pm0.39$&$0.17\pm0.53$&$0.71\pm0.28$&$0.62\pm0.51$&$0.75\pm0.68$&$-0.59\pm1.11$\\
$W$&$0.04\pm0.34$&$-0.01\pm0.46$&$1.04\pm0.25$&$0.94\pm0.45$&$0.17\pm0.60$&$-0.95\pm0.97$\\
\hline
\hline
&\multicolumn{6}{c}{\bf{Bennett et al. (constrained:$\beta_s = 2.7$; $\beta_{ff} = 2.15$)}}\\
\hline
$K$  &$-$ &&$  -  $&&$  - $&\\
$Ka$ &$-$ &&$  -  $&&$  - $&\\
$Q$  &$1.01$ &&$  1.04  $&&$ (1.92)  $&\\
$V $ &$0.34$ &&$  0.62  $&&$ (0.82)  $&\\
$W $ &$0.11$ &&$  0.87  $&&$ (0.32)  $&\\
\hline
&$KQ85$&$KQ75$& $KQ85$&$KQ75$&$KQ85$&$KQ75$\\
\hline
&\multicolumn{6}{c}{\bf{{\fastica} - function p }}\\
\hline
$K$&$6.57\pm0.58$&$6.05\pm1.09$&$5.62\pm0.29$&$4.78\pm0.51$&$8.32\pm0.72$&$8.14\pm1.15$\\
$Ka$&$2.28\pm0.57$&$2.58\pm1.08$&$2.04\pm0.28$&$1.66\pm0.51$&$4.47\pm0.71$&$4.51\pm1.14$\\
$Q$&$1.36\pm0.56$&$1.85\pm1.06$&$1.17\pm0.28$&$0.95\pm0.50$&$3.12\pm0.70$&$3.23\pm1.12$\\
$V$&$0.64\pm0.54$&$1.29\pm1.01$&$0.76\pm0.27$&$0.67\pm0.47$&$1.66\pm0.67$&$1.92\pm1.06$\\
$W$&$0.38\pm0.47$&$0.94\pm0.89$&$1.11\pm0.24$&$1.01\pm0.42$&$0.95\pm0.59$&$1.16\pm0.93$\\
\hline
&\multicolumn{6}{c}{\bf{{\fastica} - function g }}\\
\hline
$K$&$6.45\pm0.65$&$6.34\pm1.47$&$6.00\pm0.31$&$6.37\pm0.69$&$7.72\pm0.86$&$6.18\pm1.56$\\
$Ka$&$2.25\pm0.64$&$3.02\pm1.45$&$2.09\pm0.31$&$2.39\pm0.68$&$3.85\pm0.85$&$2.93\pm1.53$\\
$Q$&$1.33\pm0.63$&$2.31\pm1.43$&$1.16\pm0.30$&$1.47\pm0.67$&$2.49\pm0.84$&$1.73\pm1.51$\\
$V$&$0.65\pm0.60$&$1.76\pm1.35$&$0.69\pm0.29$&$1.05\pm0.63$&$1.03\pm0.80$&$0.53\pm1.44$\\
$W$&$0.40\pm0.53$&$1.34\pm1.19$&$1.03\pm0.25$&$1.29\pm0.56$&$0.44\pm0.70$&$-0.01\pm1.26$\\
\hline
&\multicolumn{6}{c}{\bf{$\chi^2$ analysis}} \\
\hline
$K$&$5.91\pm0.46$&$5.33\pm0.70$&$6.36\pm0.34$&$6.67\pm0.75$&$7.06\pm1.00$&$4.99\pm1.85$\\
$Ka$&$1.86\pm0.45$&$1.63\pm0.69$&$2.29\pm0.33$&$2.47\pm0.73$&$3.25\pm0.99$&$1.88\pm1.82$\\
$Q$&$0.97\pm0.45$&$0.83\pm0.68$&$1.32\pm0.33$&$1.49\pm0.72$&$1.90\pm0.98$&$0.73\pm1.79$\\
$V$&$0.30\pm0.42$&$0.22\pm0.64$&$0.82\pm0.31$&$1.01\pm0.69$&$0.45\pm0.93$&$-0.48\pm1.70$\\
$W$&$0.09\pm0.37$&$0.03\pm0.57$&$1.13\pm0.28$&$1.27\pm0.60$&$-0.06\pm0.81$&$-0.88\pm1.50$\\
\hline
\hline
\end{tabular}
\caption{Values of the coupling coefficients in antenna
temperature units determined
between the 5-year \WMAP\ data and three foreground emission templates, 
at $1^{\circ}$ resolution. The Haslam 408~MHz map is
adopted as the synchrotron template. The {\fastica} analysis
is performed using the two non-linear functions $p$ and $g$ 
and four masks of the Galactic plane.
The corresponding results for a simple $\chi^2$ analysis
are provided for comparison. In addition, we provide
the values from the \WMAP\ 1-year \citep{bennett_etal_2003} fits
to the Q, V and W-bands performed with constraints imposed
on the synchrotron and free-free spectral indices. 
The units are $\mu$K/K for synchrotron, mK/mK
for dust and $\mu$K/R for free-free emission respectively.}
\label{table_haslam_coeff}
\end{center}
\end{table}

\begin{table}
\begin{center}
\begin{tabular}{l cccccc } 
\hline
\hline
& \multicolumn{2}{c}{\bf{synchrotron}} &  \multicolumn{2}{c}{\bf{dust}} &  \multicolumn{2}{c}{\bf{free-free}}  \\ 
\hline
&$Kp2$&$Kp0$& $Kp2$&$Kp0$&$Kp2$&$Kp0$\\
\hline
&\multicolumn{6}{c}{\bf{{\fastica} - function p }}\\
\hline
$Q$&$0.18\pm0.08$&$0.15\pm0.12$&$0.40\pm0.40$&$0.12\pm0.58$&$2.16\pm0.60$&$0.67\pm0.89$\\
$V$&$-0.00\pm0.08$&$-0.03\pm0.12$&$0.63\pm0.38$&$0.38\pm0.55$&$1.33\pm0.57$&$-0.05\pm0.84$\\
$W$&$-0.05\pm0.07$&$-0.07\pm0.10$&$1.16\pm0.34$&$0.92\pm0.48$&$0.82\pm0.50$&$-0.32\pm0.74$\\
\hline
&\multicolumn{6}{c}{\bf{{\fastica} - function g }}\\
\hline
$Q$&$0.25\pm0.10$&$0.16\pm0.17$&$0.15\pm0.47$&$0.13\pm0.80$&$1.76\pm0.79$&$0.82\pm1.17$\\
$V$&$0.06\pm0.09$&$-0.02\pm0.16$&$0.41\pm0.45$&$0.41\pm0.76$&$0.94\pm0.75$&$0.05\pm1.11$\\
$W$&$0.01\pm0.08$&$-0.06\pm0.14$&$0.96\pm0.40$&$0.93\pm0.67$&$0.50\pm0.66$&$-0.24\pm0.98$\\
\hline
&\multicolumn{6}{c}{\bf{$\chi^2$ analysis}} \\
\hline
$Q$&$0.22\pm0.09$&$0.20\pm0.12$&$0.37\pm0.41$&$0.33\pm0.68$&$1.43\pm0.85$&$0.11\pm1.21$\\
$V$&$0.03\pm0.08$&$0.01\pm0.11$&$0.59\pm0.39$&$0.53\pm0.64$&$0.76\pm0.81$&$-0.47\pm1.15$\\
$W$&$-0.01\pm0.07$&$-0.02\pm0.10$&$1.10\pm0.34$&$1.04\pm0.56$&$0.29\pm0.71$&$-0.81\pm1.01$\\
\hline
 &\multicolumn{6}{c}{\bf{Hinshaw et al. (constrained:$\beta_{ff} = 2.14$; $\beta_{s} = 2$)}} \\ 
\hline 
$Q$  &$0.23$&&$0.19$&&$0.99$&\\
$V$  &$0.05$&&$0.41$&&$0.63$&\\
$W$  &$0.00$&&$0.98$&&$0.32$&\\
\hline
\hline
&$KQ85$&$KQ75$& $KQ85$&$KQ75$&$KQ85$&$KQ75$\\
\hline
&\multicolumn{6}{c}{\bf{{\fastica} - function p }}\\
\hline
$Q$&$0.28\pm0.10$&$0.44\pm0.13$&$0.27\pm0.42$&$-0.29\pm0.66$&$2.03\pm0.84$&$1.54\pm1.16$\\
$V$&$0.10\pm0.09$&$0.25\pm0.13$&$0.51\pm0.40$&$0.03\pm0.62$&$1.25\pm0.79$&$0.89\pm1.09$\\
$W$&$0.04\pm0.08$&$0.15\pm0.11$&$1.08\pm0.35$&$0.66\pm0.55$&$0.78\pm0.70$&$0.48\pm0.96$\\
\hline
&\multicolumn{6}{c}{\bf{{\fastica} - function g }}\\
\hline
$Q$&$0.30\pm0.12$&$0.35\pm0.36$&$0.06\pm0.53$&$0.11\pm1.56$&$1.34\pm0.99$&$0.39\pm1.59$\\
$V$&$0.12\pm0.11$&$0.16\pm0.36$&$0.31\pm0.50$&$0.43\pm1.10$&$0.55\pm0.94$&$-0.24\pm1.51$\\
$W$&$0.06\pm0.10$&$0.07\pm0.31$&$0.87\pm0.44$&$1.04\pm0.97$&$0.19\pm0.83$&$-0.45\pm1.33$\\
\hline
&\multicolumn{6}{c}{\bf{$\chi^2$ analysis}} \\
\hline
$Q$&$0.23\pm0.10$&$0.21\pm0.14$&$0.43\pm0.44$&$0.66\pm0.90$&$1.05\pm1.06$&$0.07\pm1.73$\\
$V$&$0.04\pm0.09$&$0.03\pm0.13$&$0.65\pm0.42$&$0.84\pm0.86$&$0.46\pm1.00$&$-0.35\pm1.64$\\
$W$&$-0.00\pm0.08$&$-0.01\pm0.11$&$1.14\pm0.37$&$1.29\pm0.75$&$0.00\pm0.88$&$-0.79\pm1.44$\\
\hline
&\multicolumn{6}{c}{\bf{Gold et al. (constrained:$\beta_{ff} = 2.14$; $\beta_{s} = 2$)}}\\
\hline
$Q$  &$0.23$ &&$  0.19  $&&$ 0.95  $&\\
$V $ &$0.05$  &&$ 0.43  $&&$ 0.60  $&\\
$W $ &$0.00$&&$ 1.01  $&&$ 0.32  $&\\
\hline
\hline
\end{tabular}
\caption{Values of the coupling coefficients in antenna
temperature units determined
between the 5-year \WMAP\ data in the Q, V and W band and three foreground emission templates, 
at $1^{\circ}$ resolution. The K-Ka map is
adopted as the synchrotron template. The {\fastica} analysis
is performed using the two non-linear functions $p$ and $g$ 
and four masks of the Galactic plane.
The corresponding results for a simple $\chi^2$ analysis
are provided for comparison. In addition, we provide where appropriate
the values from the \WMAP\ 3-year \citep{hinshaw_etal_2007} and 
5-year \citep{gold_etal_2009} fits
to the Q,V and W-bands performed with constraints imposed
on the synchrotron and free-free spectral indices. 
The units are mK/mK for synchrotron, mK/mK
for dust and $\mu$K/R for free-free emission respectively.}
\label{table_k-ka_coeff}
\end{center}
\end{table}

We note that the results obtained using the $Kp2$ and $Kp0$ masks are
very consistent with those from our three-year analysis, with values
varying at most by $0.5~\sigma$ with the previous ones.
The frequency behavior is consistent with the theoretical expectations.
The synchrotron and free-free emissions decrease with increasing
frequency, whereas the dust coefficients are consistent with the superposition
of an anomalous component with a falling contribution until approximately
$61$ GHz where a rising contribution from thermal emission is seen.

However, consideration of the results derived with the two new masks
indicates that 
it is no longer possible to define
a trend: a larger mask does not necessarily imply lower scaling
factors, as was the case for the free-free coefficients
obtained with the $Kp2$ and the $Kp0$ masks.  
On inspection of the results obtained with the
Haslam map as synchrotron template, although the $KQ75$ mask is the
largest cut provided, we generally obtain values comparable to or higher
than those returned using the $Kp2$ and $KQ85$ masks. It is certainly
a confirmation of the fact that there exist spectral or physical variations of
the foregrounds on the sky. These are particularly significant for the
free-free emission and are presumably connected to specific
regions near the Galactic plane, which are completely removed by the
$KQ75$ mask, but retained by the $Kp2$ and $KQ85$ cuts.  

We studied the dependence of the analysis on the correction applied to
the H$\alpha$ map to correct for the dust absorption. The coupling
coefficients show differences which are not very significant, but
consistent with expectation, ie. as the absorption correction raises
the H$\alpha$ intensity, so the coefficients decrease.  What is
interesting then, is that the inconsistency between the values derived
with the $Kp0$ cut compared to the other masks still remains, implying
that it is not easily associated with variations in the dust absorption.
Thus it could be related to variations in the temperature of the
ionized gas in the medium latitude regions. In fact, using the
free-free coefficients we infer values of the electron temperature
that change with the mask and that are generally lower than the
expected value of $8000$ K. This is true in particular for the $Kp0$
mask.

The mask dependence is different for the K-Ka analysis. Each of the
foreground components shows a specific trend with the cuts. In
particular, the dust and free-free coefficients obtained with the
$KQ75$ mask are now lower than those with the $Kp2$ and $KQ85$ masks.
However, this is not surprising, since the K-Ka map is a mix of
different emissions. It has contributions from synchrotron and
free-free, as well as from the anomalous component: none of which are
present in the Haslam map (at high latitudes -- there is a small
contribution from free-free emission in the Galactic plane).
Moreover, this can also be a possible explanation for the amplitudes
of the dust coupling coefficients in the Q-band which are low compared
to the Haslam case.
Specifically, the anomalous component is now traced largely by the 
K-Ka template, rather than the dust. Indeed, this may also suggest
that the thermal dust template is not morphologically identical to the
anomalous dust emission, although well-correlated.

We have also used a simple $\chi^2$ analysis as a standard method to
be compared with {\fastica}.  Whereas with the \WMAP\ 3-year data we
considered only the results for the $Kp2$ mask, here we extend the
comparison to all of the sky coverages. This choice was motivated by
the unusual dependence on the extension of the mask observed in our
analysis.  With the Haslam map as synchrotron template, the anomaly is
confirmed by the $\chi^2$ analysis.  In fact the {\fastica} numbers
are generally in good statistical agreement with the $\chi^2$ results,
even though the synchrotron and dust amplitudes are systematically
lower (higher) for the $\chi^2$ method. However, this in general
reflects the weak cross-talk seen between the fitted amplitudes.
Alternately, when the K-Ka map is adopted instead, the $\chi^2$
coupling coefficients of the synchrotron and dust emission are stable
with respect to the mask used. Conversely, the free-free coefficients
are found to be strongly dependent on the cut. Moreover, the values
obtained with the $Kp0$ and $KQ75$ masks are consistent with zero.
Nevertheless, all the coefficients are consistent with the
\fastica\ results.

Finally, we compared our results with those obtained by
\citet{bennett_etal_2003}, \citet{hinshaw_etal_2007} and
\citet{gold_etal_2009} with constrained fits.  The values are
comparable for the synchrotron and dust emissions, with both the
synchrotron templates.  The free-free values, however, are notably
different in all the cases, 
in part due to the fact that they corrected the H$\alpha$ map for dust
absorption. 
However, even if we repeat our analysis using the \WMAP\ value for the dust
absorption correction, the coefficients remain larger than \WMAP\ derive.
Therefore, it cannot be the only explanation for the difference.

\subsection{Spectral index of foreground emissions.}
\label{spectral_index}

We parameterise the spectral behavior of the synchrotron emission
(as traced by the Haslam template), the
anomalous dust component, and free-free emission with particular
emphasis on the latter. The \WMAP\ frequency range does not allow a
detailed study of the spectral behaviour of the thermal component of
dust.
We fitted the coefficients of each component with a power law model of
the form $A_{norm}(\nu/\nu_{0})^{-\beta}$. $A_{norm}$ is the amplitude
of the emission of a specific physical component at the reference
frequency $\nu_{0}$, which we take as the K-band ($23~\rm{GHz})$.  In
the case of the anomalous component, in order to isolate its
contribution we recomputed the coefficients from the sky maps after
correcting them for a thermal dust contribution assuming the FDS8 dust
model. This assumption is consistent with the W-band correlation
results in table~\ref{table_haslam_coeff}.

\begin{table}
\begin{center}
\begin{tabular}{l | c| c| c| c| c| c }
\hline
\hline
&\multicolumn{2}{c}{\bf{Synchrotron}}&\multicolumn{2}{c}{\bf{Anomalous Component}}&\multicolumn{2}{c}{\bf{Free-Free}}\\
\hline
&\multicolumn{6}{c}{{\fastica} -- function p}\\
\hline
&  $Kp2$&  $Kp0$ &  $Kp2$&  $Kp0$ &$Kp2$&  $Kp0$  \\
\hline
$\beta$  &$3.25_{-0.49}^{+1.18}$&$3.21_{-0.54}^{+1.23}$&$3.32_{-0.26}^{+0.48}$&$4.72_{-0.36}^{+0.31}$&$1.88_{-0.50}^{+0.62}$&$2.89_{-0.38}^{+1.28}$\\
$A_{norm}$&$6.20\pm0.47$&$5.46\pm0.72$&$5.79\pm0.20$&$4.87\pm0.28$&$8.57\pm0.44$&$6.76\pm0.69$\\
\hline
&  $KQ85$&  $KQ75$ &  $KQ85$&  $KQ75$ &$KQ85$&  $KQ75$ \\
\hline
$\beta$   &$2.70_{-0.46}^{+1.11}$&$1.80_{-0.75}^{+0.90}$&$3.09_{-0.33}^{+0.39}$&$3.29_{-0.50}^{+0.25}$&$1.65_{-0.58}^{+0.23}$&$1.51_{-0.41}^{+0.41}$\\
$A_{norm}$&$6.52\pm0.56$&$5.84\pm1.02$&$5.52\pm0.20$&$4.67\pm0.37$&$8.27\pm0.66$&$8.05\pm0.91$\\
\hline
&\multicolumn{6}{c}{{\fastica} -- function g}\\
\hline
&  $Kp2$&  $Kp0$ &  $Kp2$&  $Kp0$ &$Kp2$&  $Kp0$  \\
\hline
$\beta$  &$3.06_{-0.56}^{+1.35}$&$2.76_{-0.75}^{+1.07}$&$3.35_{-0.31}^{+0.42}$&$4.09_{-0.38}^{+0.54}$&$1.94_{-0.52}^{+0.31}$&$2.74_{-0.24}^{+1.28}$\\
$A_{norm}$&$6.41\pm0.55$&$5.59\pm0.90$&$6.05\pm0.21$&$5.68\pm0.35$&$8.50\pm0.60$&$6.62\pm1.09$\\
\hline
&  $KQ85$&  $KQ75$ &  $KQ85$&  $KQ75$ &$KQ85$&  $KQ75$ \\
\hline
$\beta$  &$2.69_{-0.46}^{+1.11}$&$1.41_{-0.75}^{+0.90}$&$3.23_{-0.28}^{+0.45}$&$2.79_{-0.16}^{+0.71}$&$1.99_{-0.50}^{+0.30}$&$2.28_{-0.13}^{+0.72}$\\
$A_{norm}$&$6.40\pm0.63$&$6.00\pm1.34$&$5.90\pm0.22$&$6.23\pm0.51$&$7.75\pm0.81$&$6.25\pm1.03$\\
\hline
&\multicolumn{6}{c}{$\chi^2$}\\
\hline
&  $Kp2$&  $Kp0$ &  $Kp2$&  $Kp0$ &$Kp2$&  $Kp0$  \\
\hline
$\beta$   &$3.26_{-0.48}^{+1.39}$&$3.37_{-0.54}^{+1.23}$&$3.20_{-0.38}^{+0.36}$&$3.39_{-0.32}^{+0.34}$&$2.20_{-0.55}^{+0.66}$&$3.44_{-0.24}^{+0.72}$\\
$A_{norm}$ &$5.98\pm0.41$ &$5.43\pm0.54$&$6.34\pm0.22$&$6.25\pm0.40$&$7.99\pm0.68$&$5.78\pm1.01$\\
\hline
&  $KQ85$&  $KQ75$ &  $KQ85$&  $KQ75$ &$KQ85$&  $KQ75$ \\
\hline
$\beta$  &$3.14_{-0.46}^{+1.11}$&$3.26_{-0.86}^{+0.86}$&$3.06_{-0.45}^{+0.30}$&$2.86_{-0.16}^{+0.71}$&$2.38_{-0.50}^{+0.30}$&$-3.39_{-0.41}^{+0.41}$\\
$A_{norm}$&$5.90\pm0.44$&$5.33\pm0.61$&$6.26\pm0.24$&$6.55\pm0.55$&$7.14\pm0.89$&$5.07\pm1.36$\\
\hline
\hline
\end{tabular}
\caption{Spectral index $\beta$ and normalisation
factor $A_{norm}$ obtained fitting values of the coupling coefficients for
synchrotron (as traced by the Haslam template), the anomalous component of dust and free-free emission, 
with different masks. The normalization factor $A_{norm}$ has units equal to mK/mK for synchrotron, mK/mK
for dust and $\mu$K/R for free-free emission respectively. }
\label{table_beta_index}
\end{center}
\end{table}

Since, the $Kp2$ and $Kp0$ results do not show any significant
difference with respect to those computed with the \WMAP\ 3-year data,
we do not expect the spectral index to change and this is indeed seen
to be the case in table~\ref{table_beta_index} . For the synchrotron emission, using the $Kp2$ mask the spectral
index is steeper than $\beta_s=3.0$, even though still consistent,
while it is flatter for the $Kp0$ mask, particularly when using the
$g$-function.  The values of the spectral index $\beta_a$ of the
anomalous dust component are larger than 3 especially if the $Kp0$
mask is adopted. Therefore, they are steeper than the value of $2.85$
obtained by \citet{davies_etal_2006}, although higher values of the
spectral index are expected if we consider regions of the sky at mid-
to high-latitude.  In fact, \citet{davies_etal_2006} have also noted
spectral indices as steep as 3.8 in several dust dominated regions.

However, the values obtained with the new masks highlight some
unexpected physical properties of the foreground components.  In
fact, with the $KQ85$ and $KQ75$ masks, the synchrotron and anomalous
dust spectral indices are flatter than those derived from the older $Kp2$ and $Kp0$ masks,
and in some cases considerably so. 
Indeed, whilst all values are consistent within the errors
(derived from simulations), we are unable to
explain these differences given that the new masks can be
considered as small modifications
to the older ones. The major difference relates to the omission now
of various features associated with the free-free emission. That 
such apparently small changes can affect the \fastica\ analysis to such an extent  
might be considered problematic, although it may ultimately be telling us something about
the properties of the foregrounds close to the mask boundaries.

With regards to the free-free emission, the previous analysis in
\citet{bottino_etal_2008} found suggestions of anomalous behaviour
that remain to be explained. Our fits for all of the sky coverages are
statistically consistent with the expected scaling for free-free
emission with frequency, $\beta_{ff}=2.14$. Nonetheless, depending on
the extension of the cut the spectral behaviour of the coefficients
show different trends, as already pointed out in the earlier work.  
Comparing the spectral behaviour recovered with the different Galactic
cuts, the index is always flatter than the canonical value of $2.14$
except with the $Kp0$ mask which is steeper.  
Previously, we had interpreted this result as the consequence of different
properties of the free-free emission close to the plane (perhaps a
mixture of regions with different electron temperatures), the
presence of spinning dust in the WIM, or simply cross-talk between
different physical components that confuses the spectral analysis.


Here, we consider the second option in more detail
by adopting a model proposed by \citet{dobler_etal_2008b}.
For consistency with their analysis, we consider the template fit coefficients after conversion into 
intensity units. Practically, the 
coefficients relative to the H$\alpha$ template are described as a mix
of free-free emission ($F_0$) and radiation from the WIM ($D_0$) that
together generate the spectral bump that
\citet{dobler_etal_2008b}  observe. The final term ($C_0$) with a CMB-like spectrum 
is a consequence of the initial subtraction of an estimate
of the CMB sky before undertaking the template fitting.
Thus the derived free-free coefficients are parameterised by the
following relation:
\begin{equation}
I_{mod}=F_0\left(\frac{\nu}{23GHz}\right)^{-0.15}+D_0(DL98,v_{peak}=40GHz)+C_0\left(\frac{\nu}{23GHz}\right)^2a(\nu)
\label{model_DF}
\end{equation}
where DL98 is the WIM model of spinning dust due to
\citet{draine_etal_1998}, but shifted in
order to have the maximum emission at $40$GHz, and where $a(\nu)$ represents the conversion factor from thermodynamic to antenna
temperature at a given frequency.
In fact \citet{dobler_etal_2008b}  utilise several variants of the CMB sky 
estimated by internal linear combination (ILC) methods.
Since an ILC map will necessarily contain foreground residuals, the ILC-corrected
data contain modified amounts of the expected foreground levels,
and the $C_0$ term effectively attempts to correct for this bias.
Since we did not pre-process
the data in the same way, we do not formally need to include this term
in the analysis, but do so in the expectation that the coefficient $C_{0}$ will be
consistent with zero. This is, in fact, the case, and more
importantly, as shown in table~\ref{FDC_spectral_fits},
the $D_0$ dust coefficient is also consistent with zero --
we do not find any evidence of a spinning dust admixture with the free-free
emission.
Interestingly, the simple $\chi^2$ analysis yields a spectral index that
is very consistent
with theoretical expectations ($\sim$0.14), especially with the $Kp2$
and $KQ85$ masks. 

\begin{table}
\begin{center}
\begin{tabular}{l | c | c| c | c }
\hline
\hline
&\multicolumn{4}{c}{\bf{Free-Free}}\\
\hline
&\multicolumn{4}{c}{{\fastica} -- function p}\\
\hline
&  $Kp2$&  $Kp0$ &  $KQ85$&  $KQ75$ \\
\hline
$F_0$&$8.21\pm0.63$&$7.73\pm1.01$&$7.52\pm0.97$&$7.05\pm1.55$\\
$D_0$&$0.05\pm0.10$&$0.05\pm0.16$&$0.05\pm0.15$&$0.04\pm0.24$\\
$C_0$&$0.31\pm0.38$&$-1.13\pm0.61$&$0.75\pm0.59$&$1.07\pm0.94$\\
\hline
&\multicolumn{4}{c}{{\fastica} -- function g}\\
\hline
&  $Kp2$&  $Kp0$ &  $KQ85$&  $KQ75$ \\
\hline
$F_0$&$8.24\pm0.87$&$7.43\pm1.30$&$7.59\pm1.16$&$6.49\pm2.09$\\
$D_0$&$0.05\pm0.13$&$0.06\pm0.20$&$0.05\pm0.18$&$0.06\pm0.32$\\
$C_0$&$0.19\pm0.53$&$-0.99\pm0.79$&$0.08\pm0.70$&$-0.37\pm1.26$\\
\hline
&\multicolumn{4}{c}{{$\chi^2$}}\\
\hline
&  $Kp2$&  $Kp0$ &  $KQ85$&  $KQ75$ \\
\hline
$F_0$&$8.16\pm1.00$&$7.24\pm1.62$&$7.54\pm1.35$&$6.37\pm2.48$\\
$D_0$&$0.06\pm0.15$&$0.07\pm0.25$&$0.06\pm0.21$&$0.07\pm0.38$\\
$C_0$&$-0.29\pm0.60$&$-1.65\pm0.98$&$-0.54\pm0.82$&$-1.47\pm1.50$\\
\hline
\hline
\end{tabular}
\caption{ Values of the parameters obtained fitting the free-free coupling coefficients 
with the model proposed by \citet{dobler_etal_2008b} (see equation \ref{model_DF}).
The coefficients are those derived using both {\fastica} and the $\chi^2$ analysis.
$C_0$ and $D_0$ are consistent with zero, the later result demonstrating that we 
do not find any evidence of a spinning dust admixture with the free-free
emission. The units of the parameters are kJy/sr/R.}
\label{FDC_spectral_fits}
\end{center}
\end{table}

As noted before, the different treatment of the CMB
component in the derivation of the coefficients, together with the
different fitting methodologies, seem to play important roles in attempting a
reasonable interpretation of the results and a connection to what has
been found by other authors.  Indeed, cross-talk between the CMB and foreground
emission -- the so-called \lq cosmic covariance' of \citet{chiang_etal_2009} --  can likely 
both flatten and steepen the 
determined spectrum in some cases,  and \citet{dobler_etal_2008b}
have proposed that this effect makes it mandatory to subtract a CMB estimate before
attempting any template fitting.  In fact, we have observed
such spectral changes in a set of simulations in which the foregrounds are
described with idealised spectra (ie. $\beta_s=3.1$,
$\beta_{ff}=2.15$, $\beta_d=1.7$, 
and a contribution from an H$\alpha$ correlated
WIM spinning dust component is either included or otherwise. However,
the typical behaviour is such that the latter component is not
detected erroneously when it is absent, and is detected when present.
We do not then find a likely explanation of the inconsistency of our results
with \citet{dobler_etal_2008b} given our unbiased recovery of the
input coefficients.

\begin{figure}
\begin{centering}
\includegraphics[angle=90,width=1\textwidth]{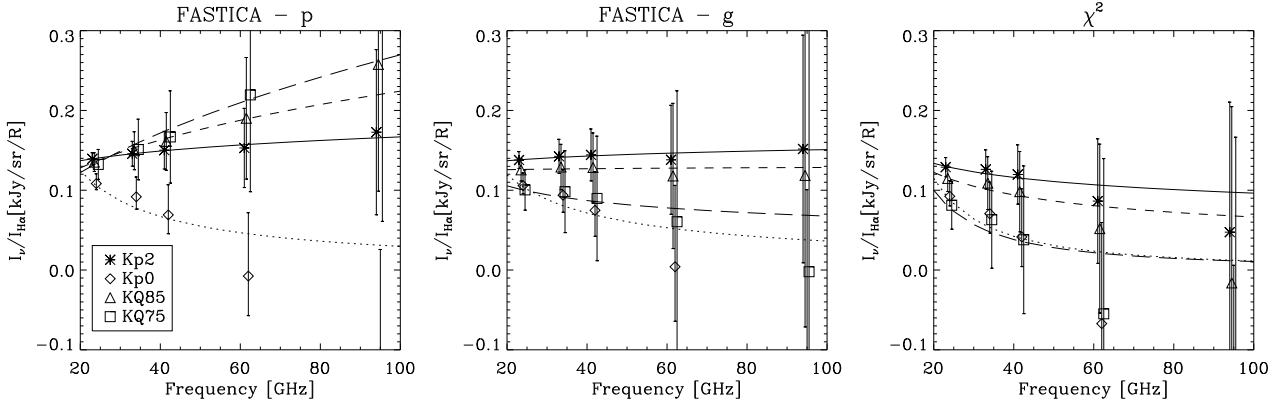}
\caption{ The coupling coefficients in intensity units (kJy/sr/R) for 
free-free emission as traced by the H$\alpha$ template. We show the results from the {\fastica} 
$p$ function analysis are in the leftmost plot, from the $g$ function
in the middle plot, and the values from the $\chi^2$ analysis are
shown on the right. Asterisks represent the derived amplitudes for the $Kp2$ sky 
coverage, diamonds are for $Kp0$, the triangles for $KQ85$ and the squares for $KQ75$. Best-fit 
curves are also shown.
The {\fastica} $p$ results show an anomalous rising
spectrum for the free-free emissivity with all the masks but the $Kp0$, whereas with the $g$ 
function and the $\chi^2$ analysis the results are more consistent with expectations.
The $Kp0$ mask shows consistently a steeper  spectrum than expected.
The $\chi^2$ results are consistent with the expectations for
free-free emission.
}
\label{beta_plot_free-free_nch_5_fd00}
\end{centering}
\end{figure}

\section{EVALUATION OF THE \WMAP\ SKY MAPS AFTER FOREGROUND CLEANING}
\label{cleaned_data}

Computing the coupling coefficients between the \WMAP\ data and the 
foreground templates is useful both to study
the spectral properties of the foregrounds and to allow the
data to be cleaned for subsequent cosmological analysis.
In our approach, we apply the \fastica\ method to large sky areas, thus
effectively assuming that each foreground has a single spectral
index over the region of interest. In reality, this is incorrect,
and we therefore expect that the cleaned data will contain residual
foreground contamination, as a consequence both of this assumption,
and the related one that the templates do indeed provide an
adequate representation  of the foreground morphology at microwave
wavelengths. The study of such residuals is then itself important
in order to gain new physical insight, and to evaluate their impact
on the statistical properties of the cleaned CMB maps. We have 
followed this approach in the 3-year analysis of \citet{bottino_etal_2008}.

In this paper, we wish to experiment further with the implementation of the cleaning
approach applied to the data. To that end, we should recall that the coefficients
are derived by maximising an approximation to the neg-entropy over the sky
fraction of interest. Such a non-linear function could be dominated
by one or two regions, that are either bright or have complex morphologies,
thus the derived coefficients would be sub-optimal elsewhere. Indeed, one
might speculate that such regions would occur close to the Galactic plane.
Conversely, with a mask that only allows high latitude regions to be analysed
where the spectral variations may be smaller, the coefficients may allow
a relatively efficient cleaning of the data and small amounts of residuals.
An interesting question then becomes by how much low latitude
residuals would increase if the high latitude coefficients are applied. 
We therefore determine the 
template coefficients as a function of mask, apply the
corresponding corrections to the multi-frequency data, and then inspect
the corrected maps. In section~\ref{iterative_analysis} we then
investigate the related impact on an iterative cleaning approach.
We show only the results
based on the $p$-function analysis, since there is little visual
difference relative to the $g$-function. 

\begin{figure}
\begin{centering}
\includegraphics[width=1\textwidth]{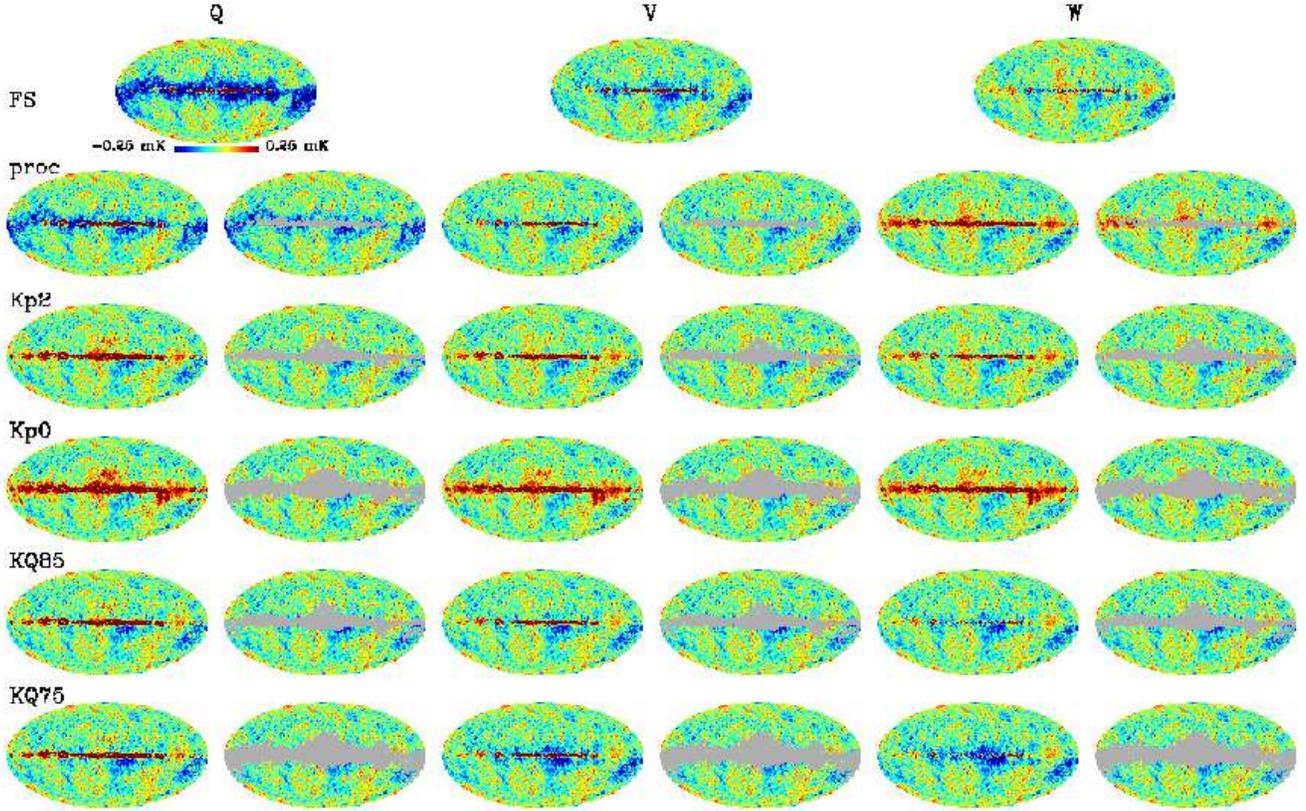}
\caption{\WMAP\ Q-, V- and W-band cleaned by
subtracting foreground templates scaled by the coupling
coefficients determined by {\fastica} with the $p$-function.
The maps are shown in a conventional mollweide
projection in a Galactic frame-of-reference , with the north pole
at the top of the image and the Galactic Center in the middle 
with longitude increasing to the left.
The regions in grey
correspond to the bright Galactic emission and point sources excised
from the analysis by the masks. 
Each row corresponds to maps cleaned by coefficients 
derived from an analysis using one of 6 masks -- FS (full sky,
although the point source mask is still employed), proc (the
processing mask), Kp2, Kp0, KQ85 and KQ75. 
The columns are divided into three pairs, each pair corresponding to
one of the three frequencies considered as indicated. The left-hand
plot in the pair is the full-sky map cleaned using the coefficients
derived when applying the indicated mask, the right-hand plot
then shows the cleaned map with the appropriate mask applied. 
The structure visible on the full-sky that falls within the
analysis mask indicates the effect of spectral mismatch between the 
low- and high-latitude sky. 
Clearly, the first row can only show one map since the two
cases coincide for a full-sky analysis.
Here, the adopted synchrotron template is the Haslam map. 
}
\label{maps_cleaned_coeff_haslam}
\end{centering}
\end{figure}

\begin{figure}
\begin{centering}
\includegraphics[width=1\textwidth]{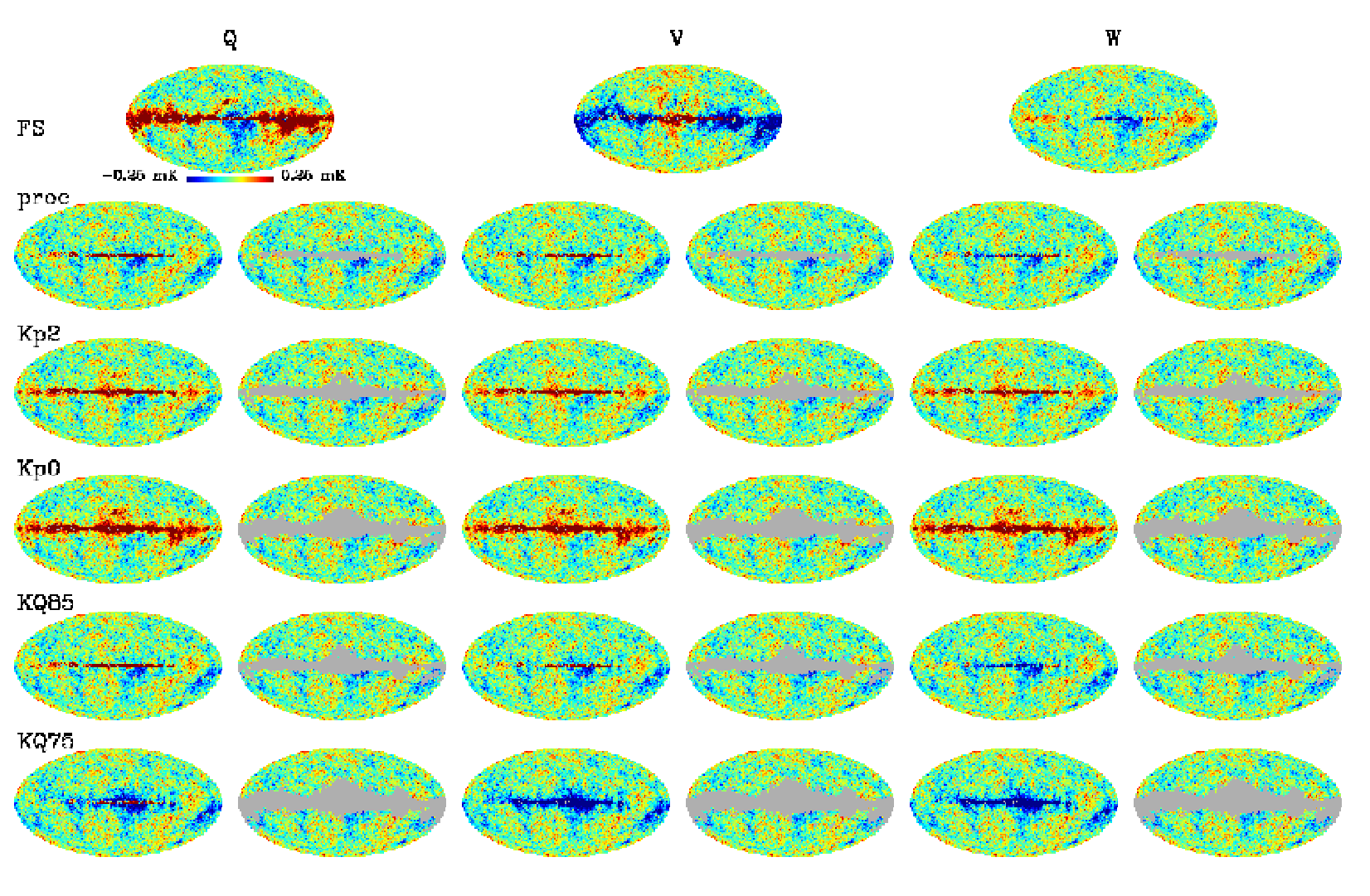}
\caption{As the previous figure, but using the K-Ka template for tracing the
    low frequency foreground emission.
}
\label{maps_cleaned_coeff_k-ka}
\end{centering}
\end{figure}

Figures~\ref{maps_cleaned_coeff_haslam} and
\ref{maps_cleaned_coeff_k-ka} show the results for the Q-, V- and
W-band data.  We show two maps at each frequency - each map
corresponds to the data as cleaned following the template analysis
with the specified mask, but both the masked and full-sky are plotted.
Showing the latter can help to reveal spectral mismatches between the
emission in the masked region and the high-latitude area used for the
analysis.  The exception is the full-sky (FS) analysis where only the
point-sources are masked in the analysis.  It is clear that there is a
complex pattern of behaviour as a function of mask, and some
dependence on the template used to trace the synchrotron
emission. However, with either the Haslam or K-Ka templates, it is
apparent that in general, the thinner the applied mask, the higher the
level of residual contamination. It should also be noted that in most
cases the $Kp0$ results seem anomalous and larger residuals are
observed.  When the Haslam template is used as the synchrotron tracer
(figure~\ref{maps_cleaned_coeff_haslam}) the Q-band residuals towards
the plane are the largest in amplitude, with the V- and W-band
comparable. However, the residuals corresponding to fits performed on
either the full-sky or using the processing mask are largely negative
for the Q-band - there is an over-correction for foregrounds towards
the Galactic plane. Interestingly, the low latitude residuals at V-
and W-band for the full-sky analysis are relatively small and positive
and confined to a very thin disk (a similar structure is seen at
Q-band though embedded in the larger region of over-subtracted
foregrounds). However, there appear to be more structures at higher
latitudes than when more extensive masks are applied. Not
surprisingly, the foreground coefficients are driven by the high
amplitude Galactic plane signal, but are then sub-optimal for the
high-latitude corrections.  For the masks commonly applied for \WMAP\
analysis, the cleaned maps at high latitude look very consistent, and
the low-latitude residuals tend to be smallest for W-band. As noted
above, the $Kp0$ results are a curiosity and the level of residuals
seems unexpectedly to be higher than for either the $Kp2$, $KQ85$ or
$KQ75$ masks, the latter case perhaps indicating some connection to
the bright free-free regions that are masked therein. This argument
may well be supported by the clear excess towards the Gum
nebula. However, with the non-linear analysis performed by \fastica\
its is difficult to be too precise about the origin of this behaviour.
For the K-Ka template, the low-latitude residuals are generally
positive except for the full-sky analysis where there is a complex mix
of positive and negative structures at Q- and V-band, and for the
$KQ75$ mask where the residuals are predominantly negative. The
results seem to suggest an improved cleaning of the data, particularly
at intermediate latitudes. It is again reassuring to note that, in all
cases, the dominant frequency-independent CMB structures are
well-pronounced at high latitudes.

\section{ITERATIVE APPLICATION OF {\fastica} ON PRE-CLEANED DATA}
\label{iterative_analysis}

In \citet{bottino_etal_2008}, we introduced the concept of an iterative application of 
\fastica\ to the data. Specifically, we applied the algorithm to sky maps
that had already been cleaned using templates with coupling coefficients
themselves derived using a \fastica\ analysis.  Here, we continue the study
begun in the previous section by applying such an iterative step
as a function of mask to data sets pre-cleaned with coefficients 
that are also a function of mask. However, we allow the mask utilised to 
determine the template coefficients to be different to that applied 
in the iterative phase to the cleaned data, and then investigate the resulting
matrix of results. We utilise either the five frequency maps cleaned using the
Haslam template, or the Q- V- and W- band data cleaned 
with either of the putative synchrotron templates (Haslam or K-Ka).

This approach to the problem is interesting in order to study
in more detail the limits of application of the algorithm.
Specifically we can address two related questions:
\begin{enumerate}
\item Can an initially poor cleaning of the data be compensated for by the
  iterative internal analysis of the cleaned data set?
\item Is it possible to derive a full-sky CMB map cleaned of any
  low-latitude residuals resulting from the application of corrections based on
  high latitude foreground coefficients?  If not, can we determine the minimal mask,  that allows
  reliable extraction of the underlying CMB signal?
\end{enumerate}
In fact, we speculate that the combination of different masks at different stages
of the iterative analysis can effectively introduce some sensitivity 
to the relative mix and spectral variations of the foreground components on the sky.

\begin{figure}
\begin{centering}
\includegraphics[width=1.\textwidth]{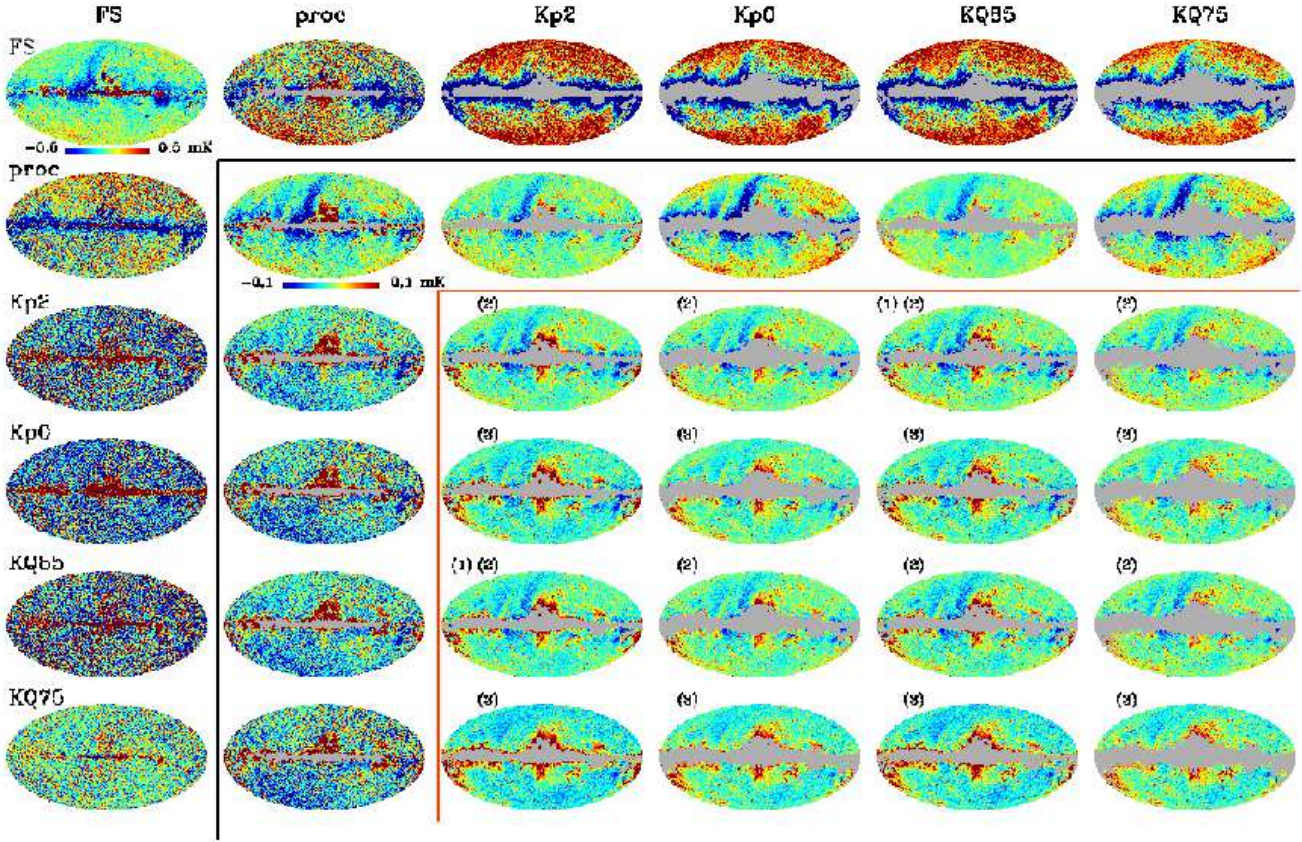}
\caption{Maps of the residual component recovered by \fastica\ with the
    iterative blind analysis of the cleaned K-, Ka-, Q-, V-, and
    W-band data. 
    Results on a given row are 
    derived from input maps cleaned using coefficients
    determined for the stated mask;
    results on a given column are derived when the stated mask is used in the
    iterative step of the analysis. The labels have the same meanings
    as figure~\ref{maps_cleaned_coeff_haslam}. 
    The numbers over the maps refer to the list of
    comments in the text.
    Here, the Haslam map
    is used as a synchrotron template. The red line highlights the cases where 
    single well-defined residual component is detected. The black line
    separates regions where different temperature scales have been
    used for the plots.
}
\label{residual_maps_masks_combination_Haslam_KKaQVW}
\end{centering}
\end{figure}

\begin{figure}
\begin{centering}
\includegraphics[width=1.\textwidth]{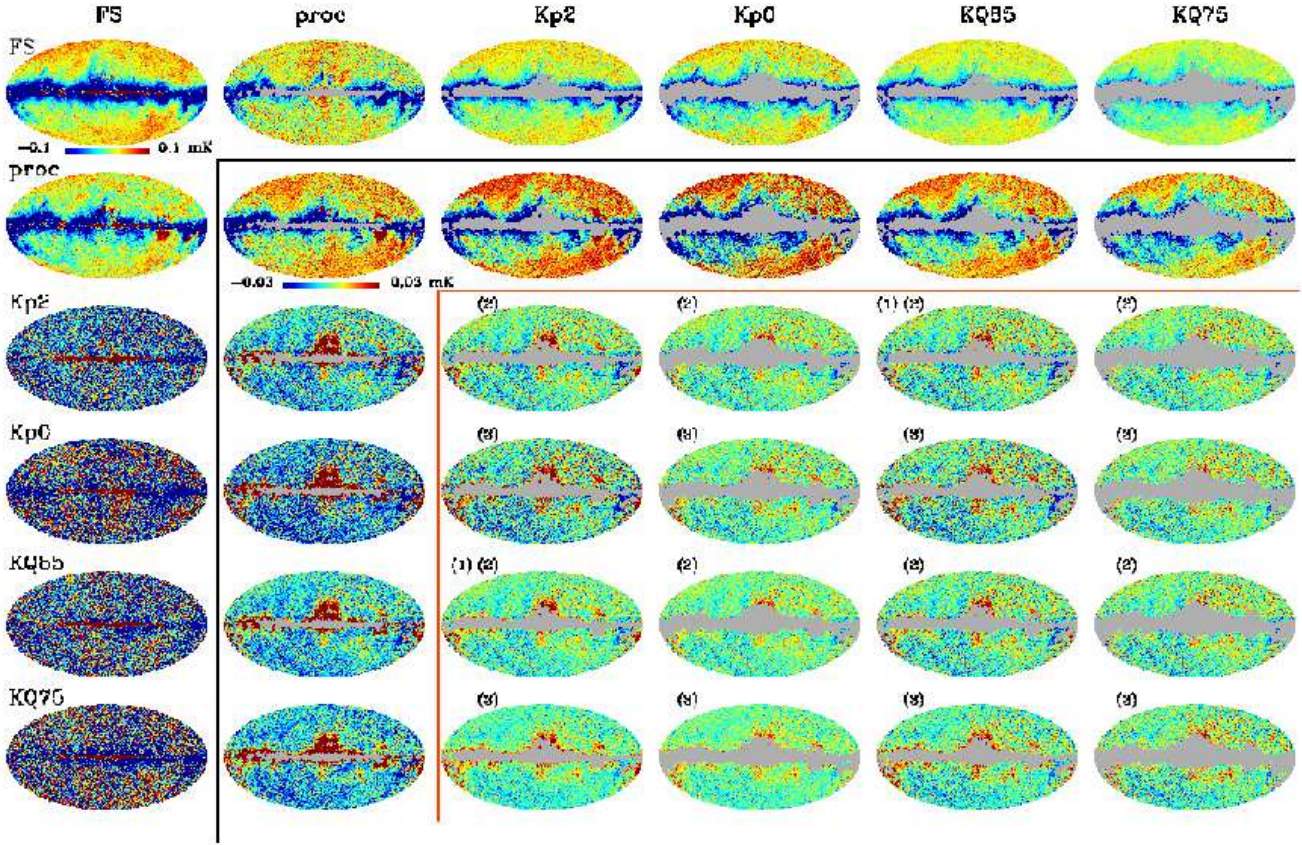}
\caption{Maps of the residual component recovered by \fastica\ with the
  iterative blind analysis of the cleaned data  Q, V and W band. Rows and columns have the same meaning as 
    figure~\ref{maps_cleaned_coeff_haslam} as well as the red box
      and the numbers, and again the Haslam map is used as the synchrotron template. 
}
\label{residual_maps_masks_combination_Haslam_QVW}
\end{centering}
\end{figure}

\begin{figure}
\begin{centering}
\includegraphics[width=1.\textwidth]{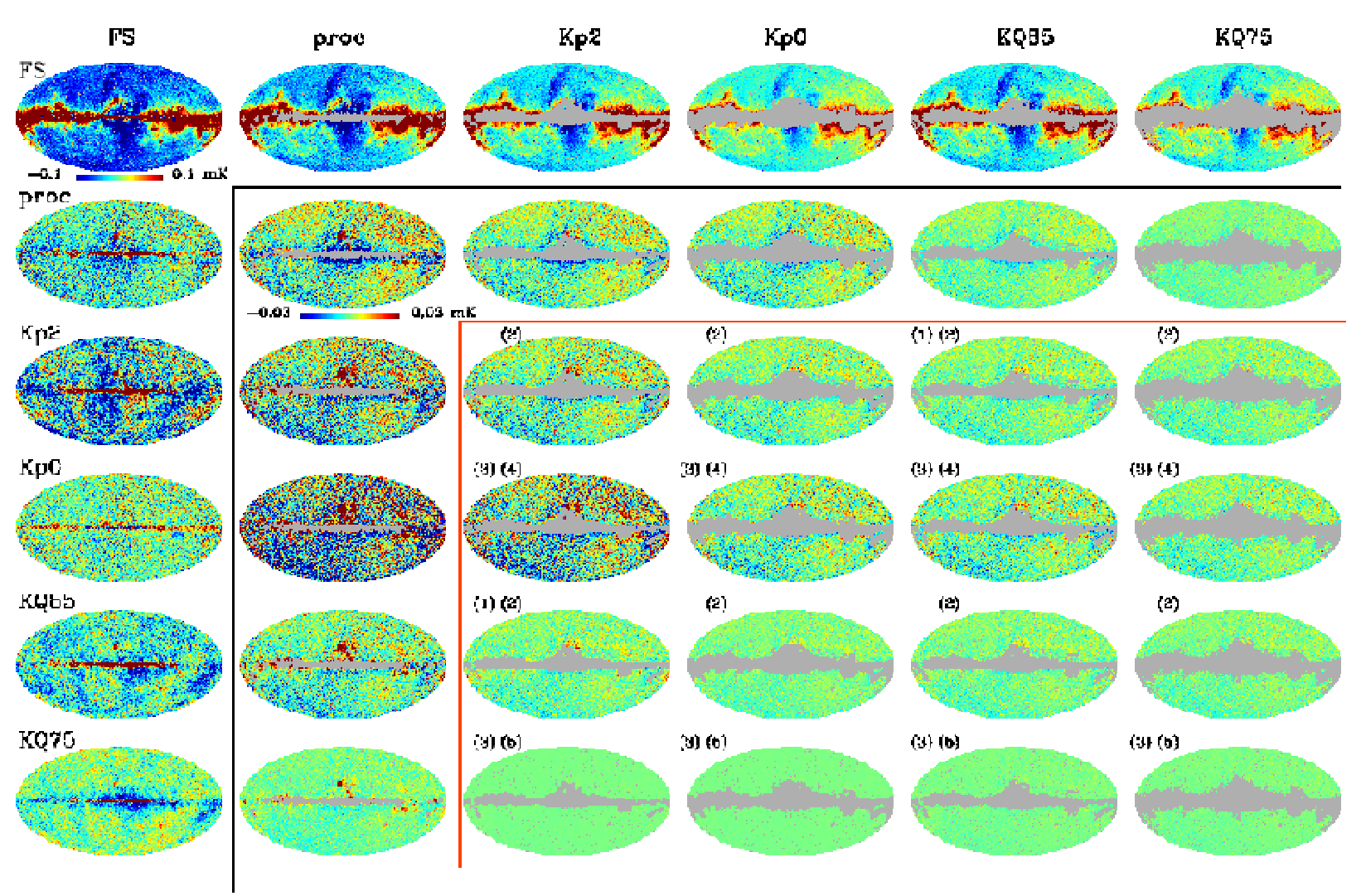}
\caption{As figure \ref{residual_maps_masks_combination_Haslam_QVW}  but using the 
    K-Ka map as the synchrotron template. 
}
\label{residual_maps_masks_combination_K-Ka_QVW}
\end{centering}
\end{figure}

In our \fastica\ implementation, for a blind analysis of $N$ input sky maps
we expect $N$ returned independent components. One of these
will be identified as the CMB sky, the remainder correspond to
other physical components on the sky. 
In figures~\ref{residual_maps_masks_combination_Haslam_KKaQVW},
\ref{residual_maps_masks_combination_Haslam_QVW}
and \ref{residual_maps_masks_combination_K-Ka_QVW}, 
we show the spatial distribution of this residual foreground
derived as a function of the mask used initially to compute the
template coefficients for cleaning the input sky maps,
and of the mask used to subsequently perform the iterative analysis.
Each row of the plot corresponds to results derived from maps
cleaned with those coefficients determined for a given mask,
whilst each column corresponds to results derived when using the stated mask
for the iterative analysis.
Consequently, the diagonal maps are those returned when the same mask is
adopted for the two steps of the analysis. 

When the standard masks (ie. $Kp2, Kp0, KQ85, KQ75$) 
are used in the two steps of the analysis, we find that only one of
the additional  non-CMB components returned by \fastica\ 
is actually consistent with a foreground residual,
the remainder being related to noise and residual dipole terms. 
A box in the plot highlights these maps which we retain as the most interesting results.
Conversely, when either the processing mask is used or a 
full-sky analysis is performed in one of the two step of the analysis,
the cleaned data remain strongly contaminated from the Galactic
foregrounds, and there is no clear single map representing foreground
residuals. Nevertheless, for completeness, we show representative maps
that contain at least part of the strong foreground signal.
In some cases, e.g. when the processing mask is used for iterative
analysis after data pre-cleaning with the standard templates, there is 
evidence of consistent residuals with those shown in the red box in
figure~\ref{residual_maps_masks_combination_Haslam_KKaQVW}. However,
it is not possible to draw strong conclusions from these cases.

If we compare the new 5-year results with the 3-year results from our previous analysis,
we can confirm the previous conclusions, which are actually reinforced
thanks to the wider set of masks adopted here.

\begin{itemize}
\item There is strong 
  evidence of a residual which is concentrated near the Galactic
  Centre and along the edge of the mask. 

 \item There is a residual associated with the North Polar Spur seen in the Haslam data, clearly visible when all five \WMAP\ 
  frequencies
  are included in the analysis and the Haslam map is used as the synchrotron template. This oversubtraction
  of the emission implies a spectral difference between this region and high latitude regions. However, the
  amplitude falls off when a larger cut is used, which may indicate some variation of this index
  along the spur's extent.

\item The spatial distribution of the contamination is larger when we analyze
  the K-, Ka-, Q-, V- and W-band data using the Haslam map as the synchrotron template
  than when only the 
  Q-, V-  and W-band maps are input data. Moreover, the amplitude of the residual is also lower
  in the latter case, implying a spectrum that falls with frequency. Whether the residual is a new
  physical component of emission or simply a region where the spectral index is flatter than
  the high latitude areas is unclear.

\item The intensity of this component gets even weaker when the K-Ka map is adopted to model the
  synchrotron emission. This is a consequence of the fact that the K- and Ka-data do contain this
  emission component which also supports the concept that the use of internal templates such
  as the K-Ka map, even with a simple global scaling, provides a more accurate means to model
  the foreground contaminations than from data obtained at much lower or higher
  frequencies. However, the physical interpretation of such templates is difficult since they
  are certainly mixtures of several components.
\end{itemize}

Including the new cases of analysis, where the masks are mixed, other interesting points come out:

\begin{itemize}

\item [1.] The processing steps with respect to masking do not commute - for example, the residuals from the iterative analysis
using a $KQ85$ mask applied to maps cleaned by $Kp2$ coefficients are not identical to the
case when the masks are reversed.

\item [2.] When the scaling factors are computed with the thinner masks (namely the $Kp2$
and the $KQ85$), the spatial distribution of the residual is similar
regardless of the mask used for the iterative analysis.
Therefore, the component is not a simple effect of the cut of the Galactic
plane.

\item [3.] The algorithm provides a good component separation also when we adopt the coefficients
relative to the largest masks: the algorithm is able to recover the residual 
whose amplitude depends on the mask applied in the second step of the analysis.
What is not subtracted in the cleaning process, is actually recovered by the
iterative analysis.

\item [4.] The residuals associated with the $Kp0$ coefficients and the K-Ka template 
seem to show unexpectedly high amplitude compared to smaller cuts and the 
similar $KQ85$ mask.

\item [5.] The $KQ75$ coefficients computed with the K-Ka template seem to be the optimal ones in order to
clean the data: the residual maps appear uniformly clean. The obvious conclusion would be 
that they give the most realistic description of the contaminated sky. However, this
idea is not always supported by analogous considerations about the CMB component. 
In fact, the set of coefficients are reliable only when the $KQ75$ mask
is used also for the iteration. In the other cases, {\fastica} mixes all the components and 
part of the residual  actually contaminates both the CMB map and a
third component, which generally shows a noise pattern.
\end{itemize}

The last point highlights that a consideration of the residuals revealed by the iterative analysis
can actually be misleading, and is not a sufficient figure of merit to quantify the performance 
of such an analysis by itself. To make such a judgement, we must consider the CMB
component returned by the analysis and, in particular, its power spectrum. However, before 
we make this assessment in section~\ref{powerspectra_iteratively_cleaned_maps}, we make a small digression.


\subsection{Is the foreground residual the \lq \WMAP\ haze'?}
\label{haze}

Since we cannot simply associate the recovered residual component with any of the 
standard templates used to describe the diffuse Galactic emissions,
we attempt to address the issue of its spatial distribution and 
physical origin. In fact, \citet{dobler_etal_2008a} have already attempted to give an answer to these
questions, and identify the residual (the so-called \WMAP\ haze) as a hard synchrotron component, whose 
origin is linked to dark matter annihilation in the centre of our Galaxy.
They also provide a simple model for its spatial distribution, namely:
\begin{equation}
\bf{h} \propto \left\{
  \begin{array}{cl}
  \frac{1}{r} - \frac{1}{r_0} & \mbox{for } r < r_0; \\
  0 & \mbox{for } r > r_0,
  \end{array}
  \right.
\end{equation}
where $r$ is the angular distance to the Galactic center and $r_0$ is arbitrarily
set equal to $45$ degrees.

\begin{figure}
\begin{centering}
\includegraphics[width=1.\textwidth]{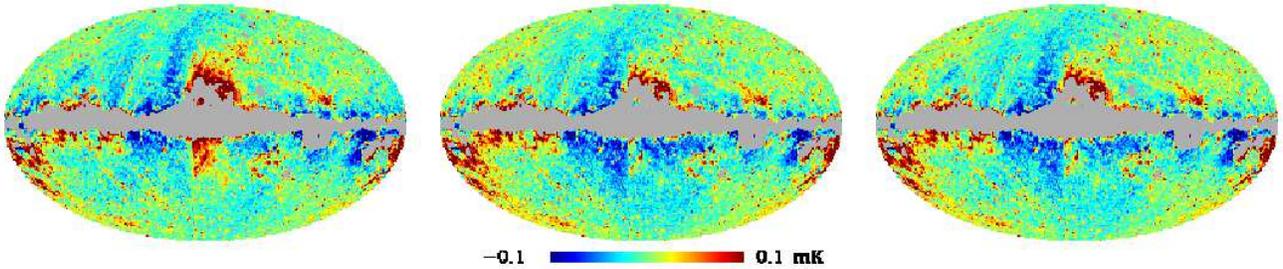}
\caption{Maps of the residual component recovered by \fastica\ with the
    iterative blind analysis of the cleaned data.  These are produced using the coupling 
    coefficients obtained from the templates fit where the haze template is not 
    used (left),  when it is added as a fourth template (center) and when the haze
    coefficients are only used to clean the the K- and Ka-band maps.
    The Haslam map is adopted as the synchrotron model. 
}
\label{residual_with_haze_as_template_new2_newcleaning}
\end{centering}
\end{figure}

Examination of the maps returned by {\fastica} indicates that a spherically symmetric
distribution is not the optimal one to describe the component. 
Nevertheless, a direct fit of this model to our maps
shows a quite good agreement, with a correlation coefficient of $0.45$. 
Therefore, we adopted it as a fourth template and repeated the iterative analysis
using the Haslam map as the soft-synchrotron template.
A physical interpretation of the coupling coefficients returned by {\fastica}
is not simple:
the values are negative for all the frequencies except the K- and Ka-bands, 
making a spectral index computation very difficult. 
Nevertheless, we used the values to clean the data
and considered the amount of contamination still present in the residual map from the
corresponding iterative analysis. 
Although these coefficients are representative of the best {\fastica} solution when the haze template is used, 
a negative value adopted to clean the data can effectivly correspond to the
introduction of a spurious foreground residual in the data.
Therefore, we also studied the case where the haze template cleaning
was only performed for the K- and Ka-bands; for the 
remaining frequencies we retained the old three-template coefficients.
Figure~\ref{residual_with_haze_as_template_new2_newcleaning} shows the
residual maps obtained in these two cases when the $Kp2$ mask is
applied to the data together with the previous result derived from the three templates fit of the data. 

When the haze template is employed, bright emission around the
Northern extension of the Galactic Centre is still present, whilst
the Southern hemisphere emission is essentially removed. 
There is little obvious difference between the two cases.
Some refinement of the haze model is still required.

\section{POWER-SPECTRUM EVALUATION OF ITERATIVELY CLEANED MAPS}
\label{powerspectra_iteratively_cleaned_maps}

Note that, in what follows we will use the \WMAP\ team's 5-year power
spectrum derived from the V- and W-band data using the MASTER
algorithm over all angular scales\footnote{Private communication:
  G. Hinshaw (2009)}
as a canonical reference. 
This should not imply that the result is indisputable, nevertheless several independent analyses
have yielded results in good agreement. Moreover, we will make comparisons on a qualitative level
regarding the broad features of the spectrum, rather than make detailed inferences about
cosmological parameters. 

\begin{figure}
\begin{centering}
  \includegraphics[angle=90,width=1.\textwidth]{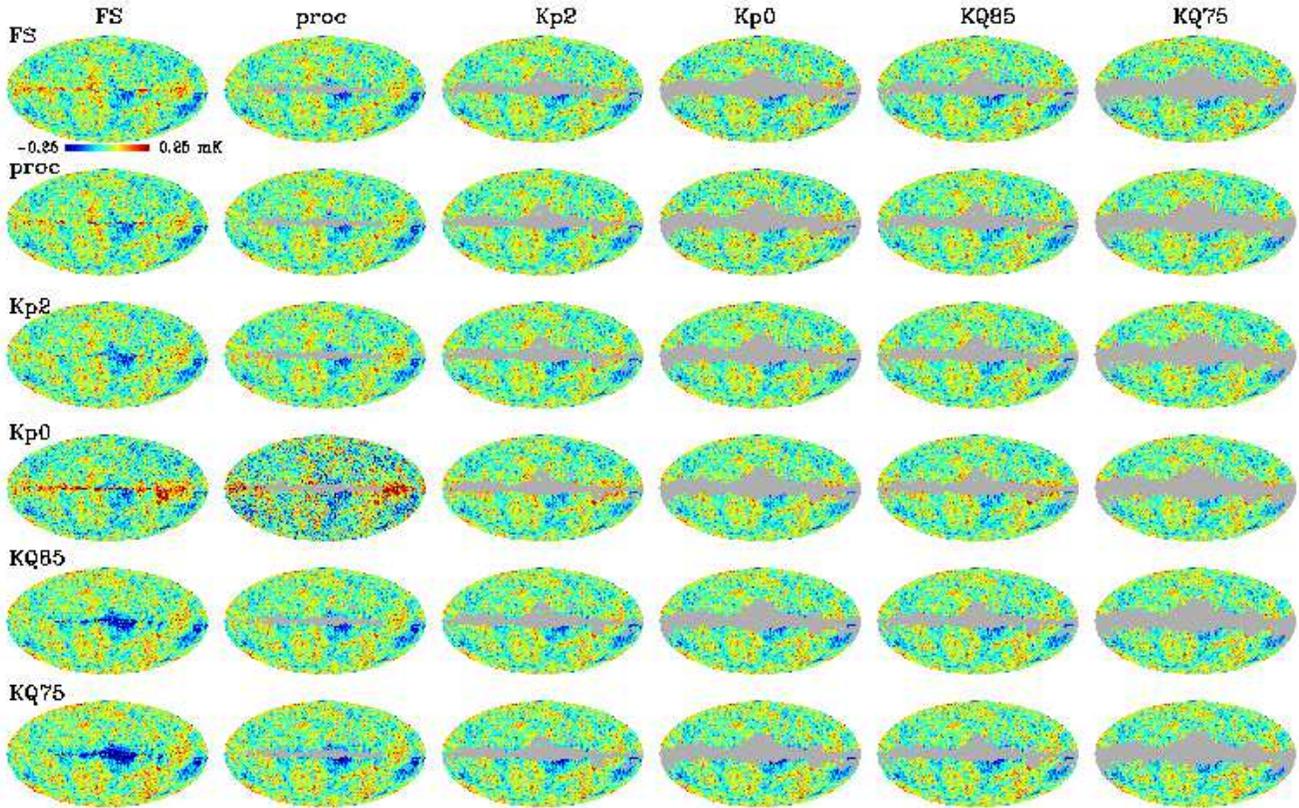}
\caption{CMB components recovered by
{\fastica} when the cleaned \WMAP\ data are used as input maps. The data
are cleaned according to the scaling factors obtained using the Haslam
map as synchrotron template, the $p$ function and \emph{all} the masks. The 
iterative analysis itself has been performed with all the different cuts 
of the sky and the five maps.
Each row is referred to the mask adopted to
compute the coefficients, further used to clean the data. Inside the row,
the single map is obtained from the internal analysis and using a specific mask, 
whose name is indicated on the top of each column.
}
\label{cmb_maps_masks_combination_Haslam_KKaQVW}
\end{centering}
\end{figure}

In figure  
\ref{cmb_maps_masks_combination_Haslam_KKaQVW},
we show the CMB maps
obtained when the five \WMAP\ bands are cleaned using the coefficients determined
from an analysis when the Haslam map is employed as the synchrotron template:
we chose this case as the best example to comment on, since it 
connects to further analyses described later on. We divide a description of the results
into two sections -- the first dealing with more conventional analysis when only partial skies are 
analysed, the second considering cases when the signal for the full sky is utilised in at least
part of the analysis.

\subsection{Partial-sky analysis}
\label{powerspectra_iteratively_cleaned_maps_cutsky}

\begin{figure}
\begin{centering}
\includegraphics[angle=90,width=0.99\textwidth]{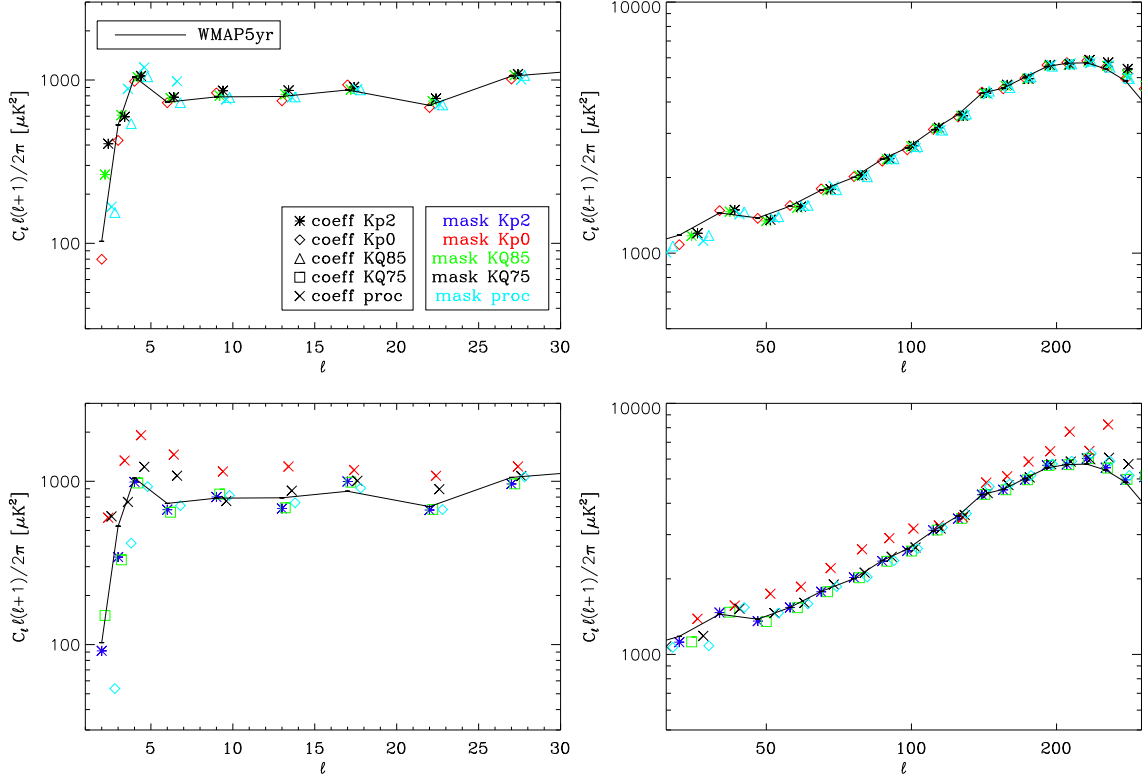}
\caption{Subset of the power spectra of the CMB components recovered by
an iterative application of {\fastica} when the cleaned \WMAP\ data are used as input maps. 
The data are cleaned according to the scaling factors obtained using the Haslam
map as synchrotron template, the $p$ function and \emph{all} the masks. The
iterative analysis itself has been performed with all the different cuts 
of the sky and the Q, V and W maps. For an input set of maps cleaned
using coefficients determined for a specific mask, best- (top row) and 
worst-case (bottom row) spectra are chosen from the corresponding
iterative analyses for all possible analysis masks using a $chi^2$
statistic relative to the best-fit WMAP spectrum.
For clarity, the left column shows the results at low-$\ell$ on a
linear scale, the right column shows the intermediate and higher-$\ell$
values on a logarithmic scale.
}
\label{plot_cl_cmb_KKaQVW_no3foreg_haslam_norm_noise_and_ps_corrected_all_masks}
\end{centering}
\end{figure}

Generally, with
all the possible combinations of coefficients and masks,
we obtain CMB maps whose power spectrum is highly consistent with
the best estimation of \WMAP\ . 
In figure~\ref{plot_cl_cmb_KKaQVW_no3foreg_haslam_norm_noise_and_ps_corrected_all_masks}
we have plotted a subset of all possible power spectra computed by
our iterative methodology. 
The selection was carried out on the basis of a reduced $\chi^2$-value 
computed relative to the \WMAP\ spectrum, and we show some
best (top panels) and worst (bottom panels) cases. 
Note that all plotted power spectra are corrected for the contributions
of the instrumental noise (determined using simulations) and the unresolved point sources
following \citet{wright_etal_2008}.

The values of the $\chi^2$-statistic are generally driven by the low
($\ell < 6$)  and high multipoles ($\ell > 250$).
In order to exclude the possibility that such excesses seen in the power
spectra could be associated with an underestimation of the noise, we implemented
a cross-power spectrum estimator. Such a method requires two input CMB sky estimates,
which we derived using subsets of the individual  DA sky maps. For Q- and V-bands, the split
is unambiguous since there are only two DAs per frequency. 
For the W band, we averaged W1 with W2 and W3 with W4. Finally, since for the K-band 
we have only one differential assembly, two independent inputs were created from averaged
maps of the first, second and third year of observations and the 
fourth and fifth year, respectively. The inverse has been done for the Ka band where the same
problem exists. The effect of these selections is to approximately balance the noise
properties of the input maps for the power spectrum evaluation.
For each of the two sets of maps, the individual frequencies were then combined with the 
coefficients derived from the standard analysis, and the cross-power spectrum evaluated.
In all cases, this has confirmed the auto-power
spectrum estimation after noise correction. 
Therefore, the excess likely reflects contamination from foreground residuals at low latitude,
arising from an inability of the \fastica\ algorithm to disentangle the components. 
This contamination is probably due to the different extension of the masks used to clean 
the maps and to internally analyse them.

However, given this interpretation, some anomalies are difficult to explain.
We would expect the contamination to be larger when the $Kp2$ and $KQ85$ masks are 
used for the internal analysis,
yet the $KQ85$ mask returns a CMB map with a lower contamination 
than for $Kp0$. 
Moreover, following this logic,
we would expect the processing mask provided by the science team of \WMAP\
to exhibit the most pronounced excess at high $\ell$, but on the contrary it is very small.
This suggests that the component separation is mostly driven by
specific features of what appears as the residual, 
rather than simply by its angular extension.
This interpretation is also appropriate to explain the results obtained from the maps cleaned using
the $Kp0$ mask coefficients and then analysed with the processing cut.
The CMB component returned by {\fastica} has a power spectrum which is consistently higher than 
the other cases, with a larger noise contribution at high multipoles.


\subsection{Full-sky analysis}
\label{powerspectra_iteratively_cleaned_maps_fullsky}

As a further experiment, we wanted to test the limits of the applicability of the algorithm
by performing a component separation study on \emph{full-sky} maps. In practice, 
by full-sky we mean the remaining sky coverage after 
exclusion of the point sources.
This experiment is interesting in order to see
if {\fastica} is able to function satisfactorily, even when the different 
sources are strongly mixed with each other, as they clearly are in the Galactic plane.

In the first stage of the analysis, we find that the component separation based on template fits is poor. 
This is only to be expected, since the spectral behaviour of the foregrounds traced by the
templates has well-established differences at low- and high-latitude, and since the bright
emission from the Galactic plane is likely to drive the fit coefficients.
We find that the coefficients are very large for the dust and free-free emissions, but small 
for the synchrotron radiation. The latter factor probably reflects the degree of mixing between the 
sources, which is particularly strong between the synchrotron and free-free emissions.
However, the general trends in spectral behaviour are maintained, 
and we adopt the coefficients to clean the data as usual. 

\begin{figure}
\begin{centering}
\includegraphics[angle=90,width=0.8\textwidth]{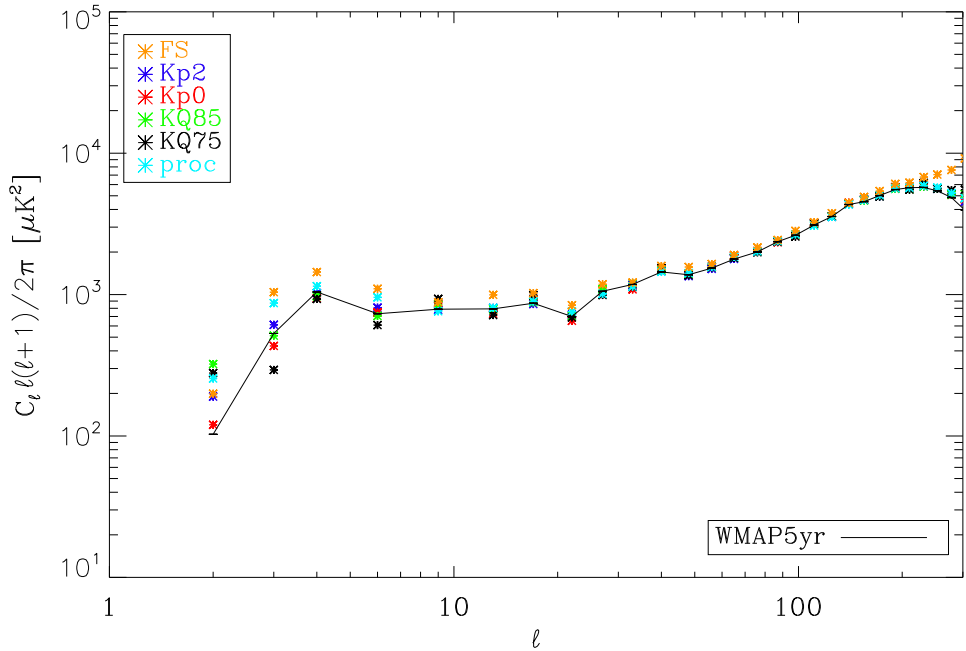}
\caption{Power Spectra of the CMB components recovered by
{\fastica} applied to the data cleaned according with the coefficients obtained with
a full-sky analysis. The Haslam map is adopted 
as synchrotron template and the coefficients were computed with the $p$ function. The
iterative analysis itself has been performed on all the five maps.
}
\label{plot_cl_cmb_KKaQVW_no3foreg_haslam_coeff_fullsky}
\end{centering}
\end{figure}

After the iterative step, the final recovered full-sky CMB map shows clearly extraneous 
features along the Galactic plane, as shown 
in the top left corner of figure~\ref{cmb_maps_masks_combination_Haslam_KKaQVW}.
The corresponding power spectrum (see figure~\ref{plot_cl_cmb_KKaQVW_no3foreg_haslam_coeff_fullsky}) 
shows, accordingly, a substantial excess
on almost all angular scales. 
Note that the enhancement on small scales ($\ell > 250$) is in the regime limited by the 1$^{\circ}$ resolution
of the analysis. Nevertheless, although we do not wish to over-emphasise its significance, it is
likely indicative of foreground residuals.

The cleaned full-sky maps (top-row of figure~\ref{cmb_maps_masks_combination_Haslam_KKaQVW}) were also iteratively analysed using the usual set of masks to exclude the Galactic plan
In these cases, we find that, on application of a mask, the excess of power at intermediate and high
multipole values largely disappears, 
confirming that most of the previously observed excess is connected to the spurious structures along the Galactic plane,
at latitudes smaller than $\sim\, 5^{\circ}$, Indeed, the processing mask itself is already sufficiently large to produce
this result.

Finally, we have internally analysed the full-sky maps cleaned utilising the 
coefficients computed using a mask (leftmost column of figure~\ref{cmb_maps_masks_combination_Haslam_KKaQVW}). 
As expected, the data appear strongly contaminated in the Galactic plane region,
since the scaling factors describe the properties of the high-latitude sky outside of a specific cut.
An intriguing result is constituted by the difference 
in the nature of this residual contamination as a function of mask:
for the high-latitude cuts, there is generally a negative region in the Galactic
plane, except when using
the $Kp0$ mask which retains positive Galactic plane structure.
This again suggests anomalous behaviour for those results derived
using this mask, following that
described in section~\ref{data_analysis} and further discussed
in section~\ref{spectral_index}, dedicated to the spectral index estimation.
The consequence of the existence of this contamination is again an excess at 
high values of $\ell$ in the CMB
power spectra,
which extends to intermediate scales most notably when using the $Kp0$
and $KQ75$ masks. None of the spectra can be considered compelling for
cosmological purposes.

\section{APPLICATION OF {\fastica} ON \WMAP\ DATA AND TEMPLATES SIMULTANEOUSLY}
\label{fastica_experiments}

From what has been presented in the previous section, we conclude that the 
experimental full-sky analysis does not satisfy our expectations.
This happens for two reasons:
first of all, we lose information on the foreground components because the template fit 
is poorly behaved; secondly, the CMB reconstruction shows significant contamination, preventing its use 
in cosmological studies.
However, we did not abandon the idea of producing an adequate full-sky CMB estimation,
and considered a data set compromising the ensemble of multi-frequency
maps dominated by the CMB component, simultaneously analysed with foreground templates.
We considered different combinations of the input data: 
the five \WMAP\  maps were analysed together with the dust and free-free maps,
and afterward also including the
Haslam map as synchrotron template. The same was done with only the Q- V- and W-band maps: 
in this case,
when the synchrotron template was involved, we used both the Haslam and K-Ka maps.
Finally, we considered the five \WMAP\ frequency channels alone, in order to 
quantify the advantages of including foregrounds models.

\begin{figure}
\begin{centering}
\includegraphics[angle=90,width=0.8\textwidth]{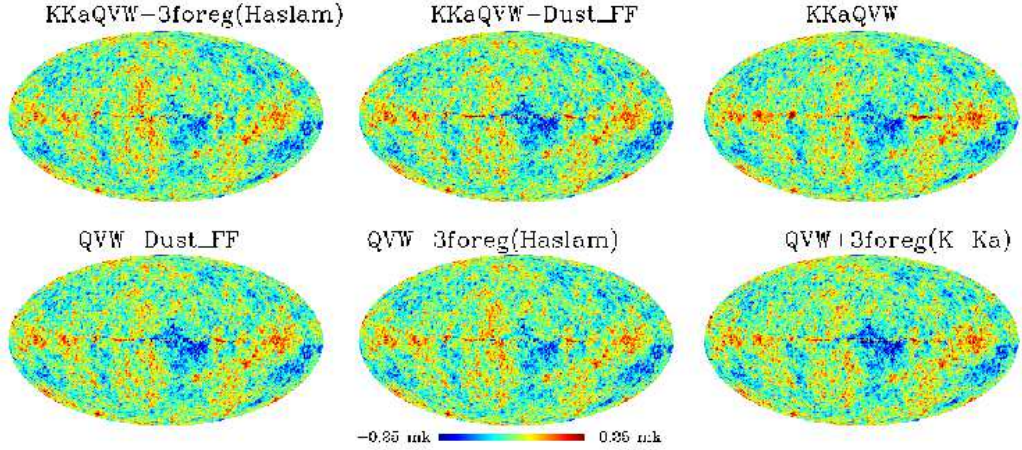}
\caption{CMB maps returned by {\fastica} when applied on different combination of data.
Several cases have been considered:
i) the \WMAP\ maps at the K-, Ka-, Q-, V- and W-bands together with three foregrounds templates, 
where the Haslam map is used to describe the synchrotron emission;
ii) the \WMAP\ maps at the K-, Ka-, Q-, V- and W-bands together with the dust 
and free-free templates;
iii) the \WMAP\ maps at the K-, Ka-, Q-, V- and W-bands;
iv) the \WMAP\ maps at the Q-, V- and W-bands with the dust and free-free templates;
v)  the \WMAP\ maps at the Q-, V- and W-bands together with three foregrounds templates, 
where the Haslam map is used to describe the synchrotron emission;
v)  the \WMAP\ maps at the Q-, V- and W-bands together with three foregrounds templates,
 where the K-Ka map is used to describe the synchrotron emission.
The cleanest map is the one obtained from the first case considered, but generally 
spurious structures are visible along the Galactic Plane in all the maps.
}
\label{cmb_maps_fastica_experiments_p}
\end{centering}
\end{figure}

\begin{figure}
\begin{centering}
\includegraphics[angle=90,width=0.8\textwidth]{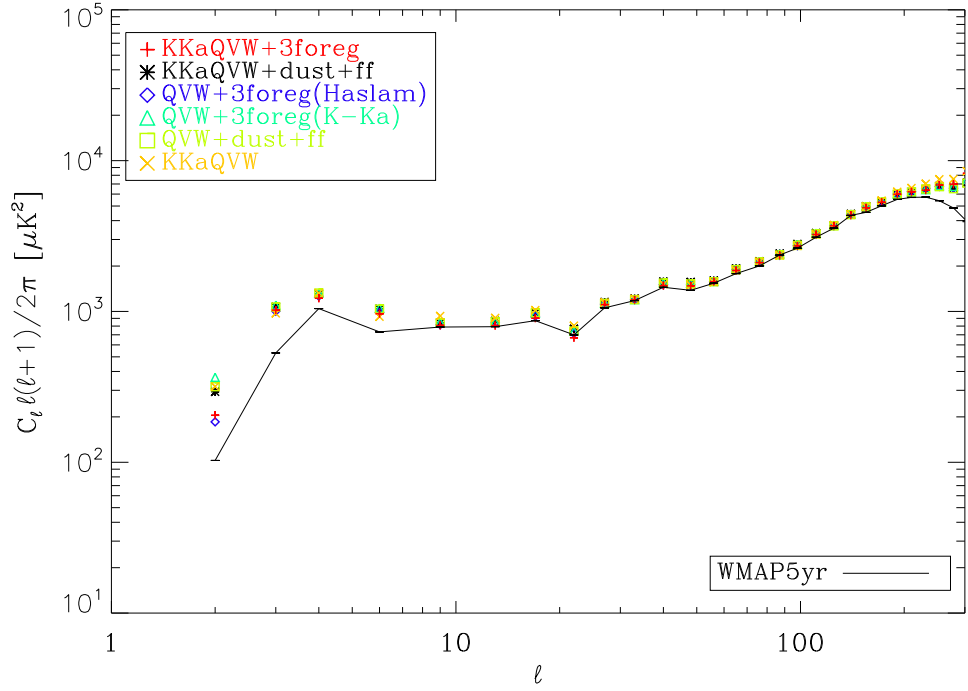}
\caption{Full-sky power spectra of the CMB maps shown in figure \ref{cmb_maps_fastica_experiments_p}. 
}
\label{plot_cl_cmb_KKaQVW+foreg_nlin_p_with_correction_ps_unresolved}
\end{centering}
\end{figure}

In figure \ref{cmb_maps_fastica_experiments_p}, we show the CMB maps obtained 
with this new implementation of the algorithm.
Unsurprisingly, it appears that the higher the number of  input maps, the greater
the precision of the CMB map reconstruction. 
In particular, with all five \WMAP\ maps and the three
foreground templates, we obtained the best CMB map with respect to those
recovered in all the other combinations of input data. Spurious structures along the 
Galactic Plane are clearly visible in all the maps, while in this case they get much weaker
and are perhaps even confined to the inner part of the Galaxy.
Quantitatively speaking, we checked the impact of these structures on the power spectrum,
shown in figure \ref{plot_cl_cmb_KKaQVW+foreg_nlin_p_with_correction_ps_unresolved}.
The CMB map returned with the largest number of input data is the most consistent 
with the best estimate from the \WMAP\ 5-year data, even though at high $\ell$ values 
there is still an excess of power, which is clearly the consequence of these residuals.

\begin{figure}
\begin{centering}
\includegraphics[angle=90,width=0.8\textwidth]{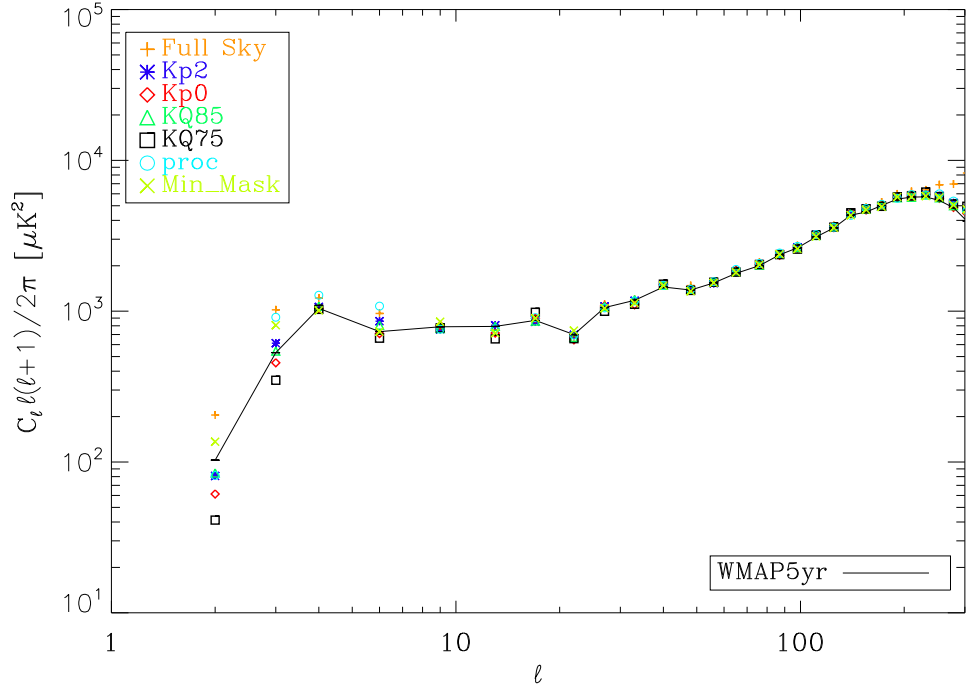}
\caption{Power spectra of the CMB map (KKaQVW+3foreg in figure~\ref{cmb_maps_fastica_experiments_p}) returned by a full-sky
    {\fastica} analysis when applied simultaneously to the five \WMAP\ maps together with three foregrounds templates. We used
    the Haslam map as synchrotron template and the $p$ function. Power-spectra are then evaluated
    after applying a variety of masks. 
}
\label{plot_cl_cmb_KKaQVW+3foreg_nlin_p_masks_noisecorrected}
\end{centering}
\end{figure}

\begin{figure}
\begin{centering}
\includegraphics[angle=90,width=0.8\textwidth]{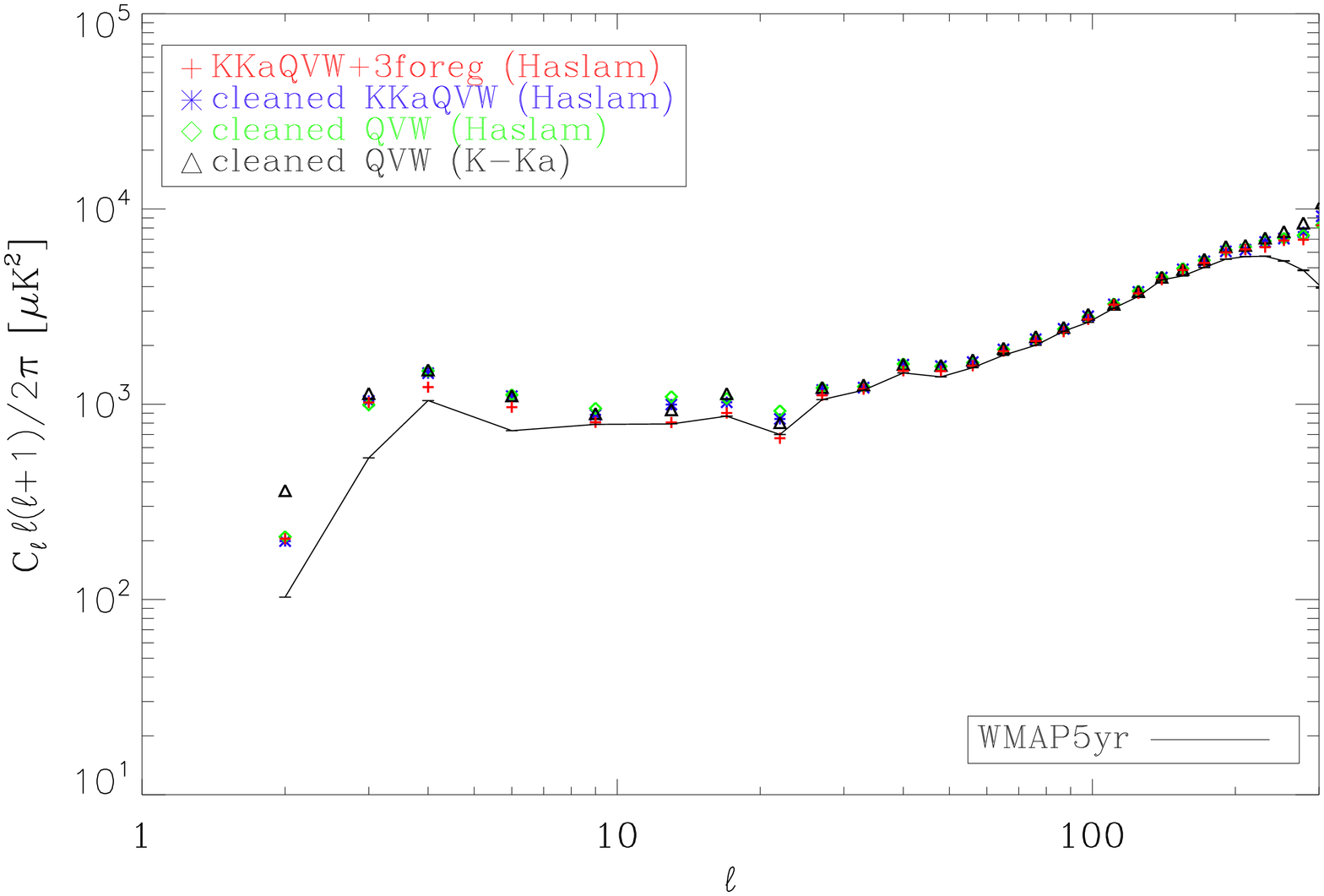}
\caption{Comparison of the  power spectrum of the CMB recovered from the simultaneous full-sky analysis
of the \WMAP\ data together with the templates and those obtained with the simultaneous internal analysis 
of the cleaned data.
We show all the different combinations of the input data internally analysed. 
The results are obtained sing the $p$ function.
}
\label{comparison_ps_KKaQVW+3foreg_vs_KKaQVWcleaned_fullsky}
\end{centering}
\end{figure}

However, we focused our attention on the map obtained from the combination
of the largest data set, computing as usual  the power spectrum as a function of applied mask.
This analysis, shown in figure~\ref{plot_cl_cmb_KKaQVW+3foreg_nlin_p_masks_noisecorrected}
seems to confirm that the excess is related to structures observed along the plane
since it disappears with more extensive cuts.
Moreover, in figure 
\ref{comparison_ps_KKaQVW+3foreg_vs_KKaQVWcleaned_fullsky} 
we compared its power spectrum to those 
of the CMB maps derived with the internal 
full-sky analysis of the cleaned data. 
The aim of this exercise was to verify that the new implementation of {\fastica}
was an improvement on previous attempts.

The excess of power for $\ell > 200$ is smaller than in the other cases
of comparison, even though the difference is not marked and we
  are at the limit of the multipoles range imposed by the
  data angular resolution of $1^{\circ}$.
On the other hand, however, at intermediate values of the multipoles, it is more consistent  
with the \WMAP\ best estimation, while the other power 
spectra are slightly higher in amplitude.

\begin{figure}
\begin{centering}
\includegraphics[angle=90,width=0.8\textwidth]{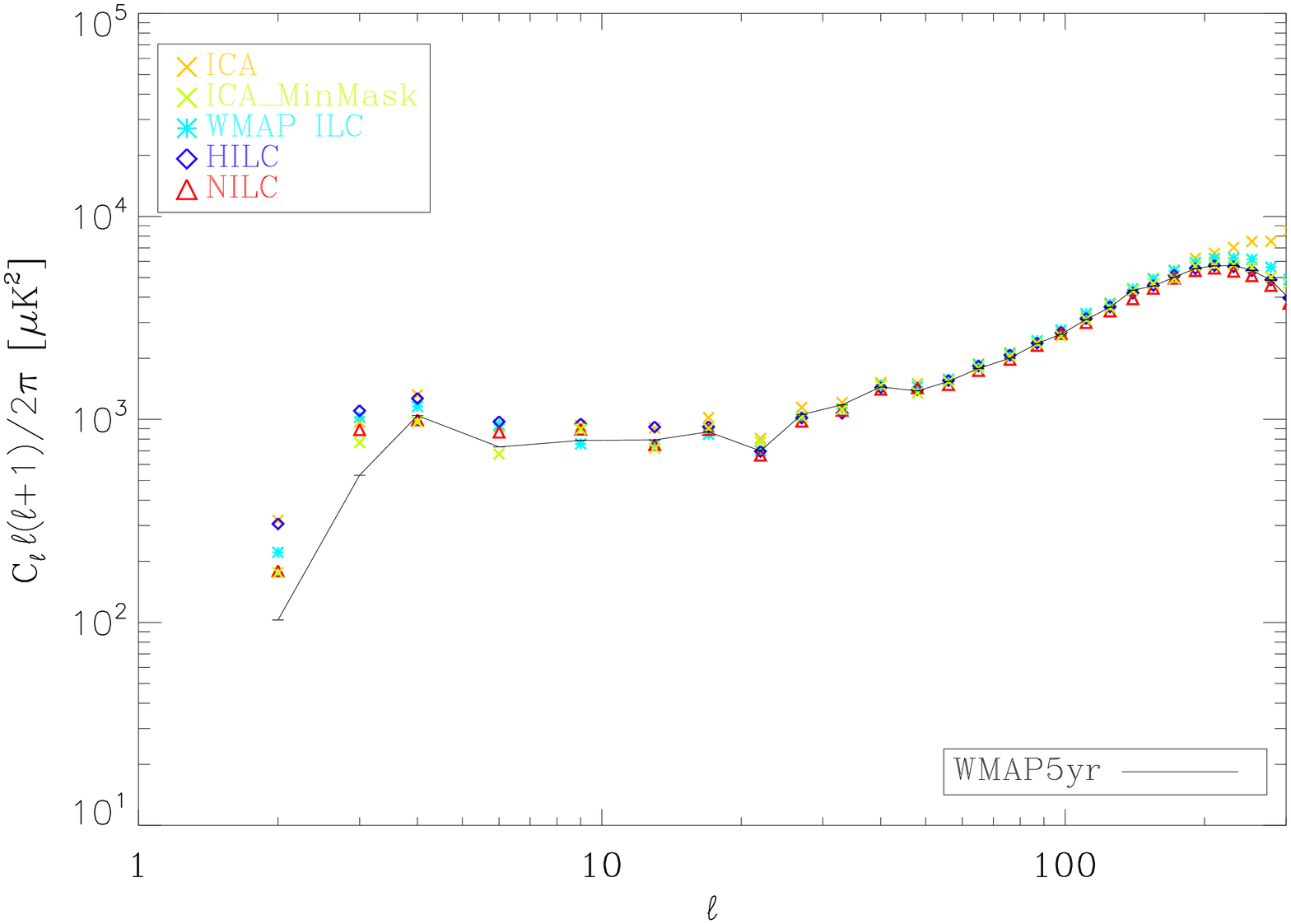}
\caption{Power spectrum of our best full-sky map obtained with {\fastica}, compared with 
those of the \WMAP\ ILC map, the HILC map produced by \citet{kim_etal_2008} and the NILC map 
of \citet{delabrouille_etal_2008}. 
}
\label{ps_ilc_maps}
\end{centering}
\end{figure}

Ultimately, we want to compare the \fastica\ method
with the results of other methods currently used to achieve the same aim.   
We considered the ILC map produced by the \WMAP\ science team \citep{gold_etal_2009}, 
a similar foreground-reduced map in which the frequency dependent
weights were determined in harmonic space by \citet{kim_etal_2008}
(hereafter HILC) and a further alternative obtained by 
\citet{delabrouille_etal_2008} using needlets as the basis of the
analysis (hereafter NILC).
Since these techniques make use of the \WMAP\ data alone without any
augmentation by internal templates, we revert to the fully blind
analysis also performed for the \WMAP\ three-year data in
\citet{maino_etal_2007}, and consider the case when all five frequency
bands are used.
Again, the power spectrum has been chosen as the figure of merit to compare the 
performances of these methods. These are shown in figure \ref{ps_ilc_maps}.
We have used published information to compute the noise correction to
the spectra based on simulations, except in the NILC case where a
correction has been directly provided, though here no correction for
unresolved point sources has been made.
The power spectra of the HILC and the NILC maps are the most consistent with the 
\WMAP\ best estimation, 
whereas the ICA map seems to be the most contaminated with residuals, 
since it shows the largest excess of power on small angular
scales. However, ins some sense, the comparison is unfair towards
\fastica\ since the other methods allow some regional dependence of
the weights used to form the optimal CMB estimate, either by solving
for coeffients in different regions of the sky separately, or finding
a best spatially varying set of weights. 

In conclusion, the ICA approach affords a fair estimate of the  CMB
signal, although improvements are still possible. However, at this
stage, we simply took advantage of this result to address another question.  

\begin{figure}
\begin{centering}
\includegraphics[angle=90,width=0.6\textwidth]{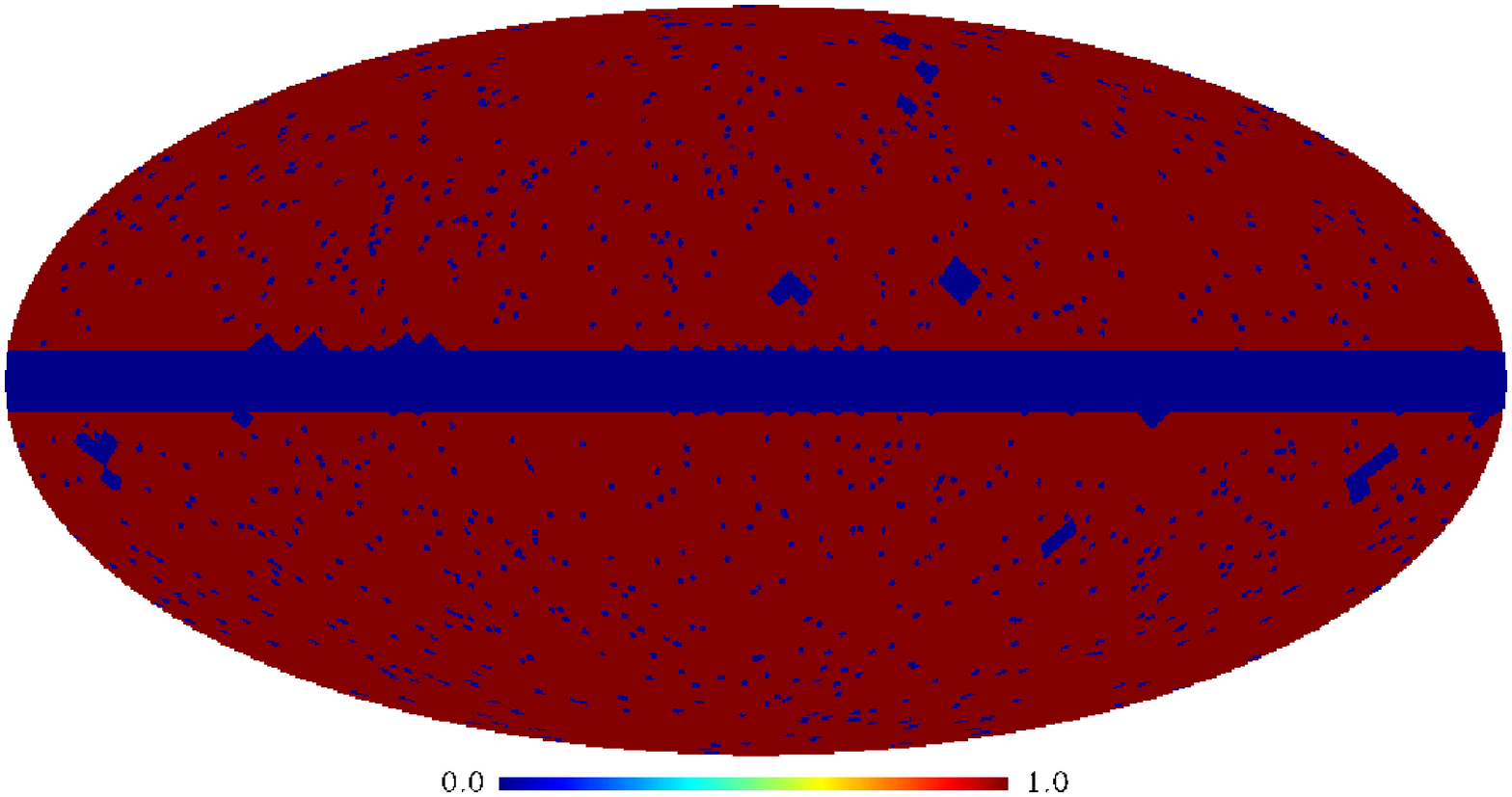}
\caption{Minimal mask derived from the {\fastica} simultaneous analysis
of the five full-sky \WMAP\ data with the three foreground templates. It is basically 
the combination of the processing mask and a parallel cut of the regions of the Galactic 
plane with $|b|<6^{\circ}$. As usual we also excluded the point sources, according with the mask
provided by the \WMAP\ science team.
}
\label{minimal_mask}
\end{centering}
\end{figure}
\subsection{What is the minimal mask?}
\label{minimal mask}
Given the imperfections of our full-sky CMB map reconstruction, 
we are forced to apply a cut in order to use it to extract cosmological information.
We therefore took the opportunity to define the minimal mask required by {\fastica}
to recovered a non-contaminated CMB map.

In order to determine the mask, we first implemented the \WMAP\ thresholding method used to generate the 
$Kp2$ and $Kp0$ masks as described by 
\citet{bennett_etal_2003}, but here applied to the {\fastica} map.
However, we found that, given the relatively low amplitudes of the residuals,
pixels for a given threshold set soon included likely genuine CMB structures, which then 
contaminated the mask . 
Consequently, we considered simple parallel cuts of different latitude extension.  
Finally, we determined a minimal mask, which is the union of the processing mask and a
parallel cut of the regions with $|b|<6^{\circ}$. As usual, 
we also excluded the point sources, according to the mask
provided by the \WMAP\ science team. The result is shown in figure \ref{minimal_mask}. 
Figures~\ref{plot_cl_cmb_KKaQVW+3foreg_nlin_p_masks_noisecorrected}
and \ref{ps_ilc_maps} then show the power spectra derived from the
\WMAP\ data either including or excluding the three foreground
templates respectively for this mask. In both cases, the agreement
with the \WMAP\ spectrum is impressive over all $\ell$, with
the remaining differences being on large angular scales.

\section{DISCUSSION}
\label{discussion}

In this paper, we have undertaken a
foreground analysis of the \WMAP\ 5-year data using the  {\fastica} algorithm, 
as previously applied to the \WMAP\ 3-year data in \citet{bottino_etal_2008}.
Various improvements in the implementation have allowed us 
to address several open questions from our previous work, and allowed
some experimentation with the technique.

We used the code to perform a foreground fit of the data
on a frequency-by-frequency basis, where the Galactic components are described by 
all-sky templates obtained at wavelengths where the corresponding emission
mechanisms dominate. Specifically, we adopted the
\citet{finkbeiner_2003} H$\alpha$-map as a template for the free-free
emission, the \citet{finkbeiner_etal_1999} FDS8 model for thermal
dust emission, and the
408~MHz radio continuum all-sky map of \citet{haslam_etal_1982} as
utilized in the first year \WMAP\ analysis for the synchrotron emission.
In the latter case, we have also considered 
the differenced of the \WMAP\ K- and Ka-band data, as preferred in their 3-year analysis.

The first step of the analysis is the computation of the
coupling coefficients, which give an estimation of the contamination
of the foreground emission level in the data.
We know already that these coefficients depend on the 
extension of the mask used to exclude the most contaminated regions
of the Galactic plane.
We further investigated this dependence, thanks to the new $KQ85$ and $KQ75$ masks provided 
by the \WMAP\ science team. We confirmed the result which reinforces
the idea of a spatial variation of the foreground emissions. Moreover, 
it suggests the component separation is driven by some key regions, rather than
the extension of the mask itself.
On the other hand, we reestablish some anomalies obtained 
when the $Kp0$ mask is adopted that remain unexplained.

We have considered the spectral behaviour of the derived scaling
factors when the \citet{haslam_etal_1982} data is used as the
synchrotron template, since an interpretation of the analogous results
derived using the K-Ka template is compromised by the fact that it
is itself a mixture of foreground components.  We evaluated the spectral index for the
synchrotron emission, the anomalous dust-correlated component, and the
free-free emission.  In the first two cases, we found steeper, though
statistically consistent, spectral behaviour as compared to previous
analysis, e.g. \citet{davies_etal_2006}. 

We then focused our attention on the estimation of the free-free spectral index.
A non-trivial dependence of the spectral behaviour on the extension 
of the mask confirms what was already found with the \WMAP\ 3-year data.
However, an understanding of the behaviour is probably associated with
two factors --
the spatial variation of the physical properties of the 
component (e.g. electron temperature which directly relates to the
scale-factor associated with the H$\alpha$ template) which  is mostly linked to specific structures
close the Galactic plane, and the connection of the statistical properties 
of the foreground with the response
of the algorithm as a function of the non-linear function used.
The spectral behaviour is generally 
flat and even increasing with frequency, if the $p$ function is used
together with the $KQ85$ and $KQ75$ masks. We have interpreted these spectral features
as a consequence of a self-similarity of the
Gaussian/non-Gaussian mixtures at each frequency.
When the $g$ function is adopted, a more uniform behaviour, consistent with theoretical expectations,
is obtained.
Finally, in both the cases, the $Kp0$ mask gives a 
steeper spectral index than the expected value of $2.14$.
We note that none of our results are compatible with the existence of a bump
in the spectrum, as claimed  by \citet{dobler_etal_2008b} and
subsequently explained by the
existence of spinning dust in the WIM that is traced morphologically by the
H$\alpha$ template.

A more extensive approach to the cleaning of the data has been introduced in connection with the iterative 
application of 
{\fastica}. A set of coefficients obtained with a specific mask is used to remove
the foreground contaminations in the maps, which are then internally analysed
with \emph{all} the Galactic cuts available.
This new approach allowed us to check the capability of the iterative step to recover 
from a poor initial cleaning of the data and still yield a credible CMB estimate.
A useful figure of merit for the quality of the results is then the corresponding 
CMB power spectrum 
as compared to the \WMAP\ estimation.
In general, the iterative approach is very robust, 
although some not completely satisfactory results provide 
a evidence of limitations of the algorithm. Specifically, it seems 
that the algorithm provides poorer estimates of the CMB
if applied on smaller regions of the sky.

Associated with the iterative analysis, we were able to determine the presence of a residual 
foreground whose spatial distribution is concentrated along the Galactic plane, with pronounced
emission near the Galactic center. 
The extension of this residual is independent of the mask used, meaning that it is not 
just the effect of a poor component separation. 
However, it decreases in amplitude if the K and Ka band are not included in 
the input data set, and more so if the K-Ka map is used to compute the synchrotron coefficients.
This indicates that the component has a falling spectrum with frequency,
but whether it is a new physical component or simply a manifestation of a region
with different spectral behaviour from the average is difficult to determine.
In any case,
it does confirm the utility of the K-Ka data as a better template
to describe the low-frequency foreground mixture.
In fact, this emission was already observed in the SMICA analysis
of \citet{patanchon_etal_2005} and is clearly consistent with the \WMAP\
haze of \citet{finkbeiner_2004}. The putative model of its spatial distribution proposed by 
\citet{dobler_etal_2008a} is found to be in reasonable agreement with the data,
although some refinements are required.

Finally, we attempted a full-sky analysis of the same data set.
Even though the code then challenged to work in regions of the sky where the cross-talk among the
components is very high, we find that {\fastica} is still able to achieve good results, particularly
based on the iterative analysis of the cleaned data as originally
implemented in \citet{bottino_etal_2008}.
We focused our attention on the CMB reconstruction, comparing the results with 
those produced by a simultaneous analysis of the multi-frequency \WMAP\ data 
with the templates.
In this case, the result is slightly better.
On the other hand, a direct comparison of the power spectra of the ICA
CMB map with variants of the ILC approach proposed recently (HILC, NILC)
suggests the need to include more spatial dependence in our analysis.
Indeed, the excess of power observed on small angular scales is likely
the signature of some residual structures along the Galactic plane. 
Consequently, to partially compensate for such structures, 
we defined a minimal mask to be used in a practical
analysis of the derived CMB map. The resulting \fastica\ power spectra
for our preferred data sets then agree remarkably with the best
estimate of the CMB spectrum provided by the \WMAP\ team.

\section*{Acknowledgments}

Some of the results in this paper have been derived using the HEALPix
\citep{gorski_etal_2005} package. We acknowledge the use of the Legacy
Archive for Microwave Background Data Analysis (LAMBDA). We thank Gary
Hinshaw for providing the best-fit \WMAP\ power spectrum derived from
the V- and W-band data using the MASTER algorithm over all angular scales. Support for
LAMBDA is provided by the NASA Office of Space Science. 
We acknowledge the use of Craig Markwardt's fitting package 
MPFIT\footnote{http://cow.physics.wisc.edu/$\sim$craigm/idl/fitting.html}.

\label{lastpage}
\end{document}